\documentclass[12pt]{article}

\usepackage{scicite}

\usepackage{times}
\usepackage{amsmath, amssymb}
\usepackage{url}
\usepackage{graphicx}
\usepackage{subcaption}
\usepackage{caption}
\usepackage{threeparttable}
\usepackage{multibib}
\usepackage{svg}
\usepackage[amssymb]{SIunits}

\usepackage{xcolor}
\usepackage{ulem}

\def\aap{Astron. \& Astrophys.}                
\newcommand{\aj}{Astron. J.}
\newcommand{\aapr}{Astron. Astrophys. Rev.}
\newcommand{\apjs}{Astrophys. J.}
\newcommand{\apj}{Astrophys. J.}
\newcommand{\apjl}{Astrophys. J.}
\newcommand{\bain}{Bull. Astron. Inst. Neth.}
\newcommand{\mnras}{Mon. Not. R. Astron. Soc.}

\newcommand{\nat}{Nature}

\newcommand{\pasp}{Pub. Astron. Soc. Pacific}
\newcommand{\physrep}{Phys. Rep.}

\newcommand{\zap}{Zeitschrift f\"ur Astrophysik}
\newcommand{\ssr}{Space Science Reviews}

\newcommand{\degr}{\hbox{$^{\circ}$}}
\newcommand{\fs}{.\!\!$^{\rm s}$}
\newcommand{\farcs}{.\!\!$^{\prime\prime}$}

\def\X-#1{X\nobreakdash-#1}

\newcites{main}{References}
\newcites{methods}{References for Methods}

\newenvironment{sciabstract}{%
\begin{quote} \bf}
{\end{quote}}

\title{Cygnus~X-1 contains a 21-solar mass black hole -- implications for massive star winds}

\author
{James C.A.\ Miller-Jones,$^{1\ast}$ Arash Bahramian,$^{1}$ Jerome A.\ Orosz,$^{2}$ Ilya Mandel,$^{3,4,5}$\\ Lijun Gou,$^{6,7}$ Thomas J.\ Maccarone,$^{8}$  Coenraad J.\ Neijssel,$^{3,4,5}$ Xueshan Zhao,$^{6,7}$\\ Janusz Zi\'{o}\l{}kowski,$^{9}$ Mark J.\ Reid,$^{10}$ Phil Uttley,$^{11}$ Xueying Zheng,$^{6,7}$\\ Do-Young Byun,$^{12,13}$, Richard Dodson,$^{14}$ Victoria Grinberg,$^{15}$ Taehyun Jung,$^{12,13}$\\ Jeong-Sook Kim,$^{12}$ Benito Marcote,$^{16}$ Sera Markoff,$^{11,17}$ Mar\'ia J.\ Rioja,$^{14,18,19}$\\ Anthony P.\ Rushton,$^{20,21}$ David M.\ Russell,$^{22}$ Gregory R.\ Sivakoff,$^{23}$\\ Alexandra J.\ Tetarenko,$^{24}$ Valeriu Tudose,$^{25}$ Joern Wilms$^{26}$}

\date{}

\begin{document} 


\baselineskip 24pt


\maketitle

\noindent
1:  International Centre for Radio Astronomy Research -- Curtin University, Perth, Western Australia 6845, Australia\\
2: Astronomy Department, San Diego State University, San Diego, CA 92182-1221, USA\\
3: School of Physics and Astronomy, Monash University, Clayton, Victoria 3800, Australia\\
4: OzGrav: The Australian Research Council Centre of Excellence for Gravitational Wave Discovery, Australia\\
5: School of Physics and Astronomy, University of Birmingham, Edgbaston, Birmingham B15 2TT, United Kingdom\\
6: Key Laboratory for Computational Astrophysics, National Astronomical Observatories, Chinese Academy of Sciences, Beijing 100012, China\\
7: University of Chinese Academy of Sciences, Beijing 100012, China\\
8: Department of Physics \& Astronomy, Texas Tech University, Lubbock, TX 79409-1051, USA\\
9: N. Copernicus Astronomical Center, PL-00-716 Warsaw, Poland\\
10: Center for Astrophysics $|$ Harvard \& Smithsonian, Cambridge, MA 02138, USA\\
11: Anton Pannekoek Institute for Astronomy, University of Amsterdam, 1098 XH Amsterdam, The Netherlands\\
12: Korea Astronomy and Space Science Institute, Daejeon 34055, Republic of Korea\\
13: University of Science \& Technology, Daejeon 34113, Republic of Korea\\
14: International Centre for Radio Astronomy Research -- The University of Western Australia, Crawley, Western Australia 6009, Australia\\
15: Institut f\"ur Astronomie und Astrophysik, Universit\"at T\"ubingen, 72076 T\"ubingen, Germany\\
16: Joint Institute for Very Long Baseline Interferometry European Research Infrastructure Consortium, 7991~PD Dwingeloo, The Netherlands\\
17: Gravitation \& Astroparticle Physics Physics Amsterdam Institute, University of Amsterdam, NL-1098 XH Amsterdam, The Netherlands\\
18: Commonwealth Scientific and Industrial Research Organisation Astronomy and Space Science, Perth, Western Australia 6102, Australia\\
19: Observatorio Astron\'omico Nacional, Instituto Geogr\'afico Nacional, 28014 Madrid, Spain\\
20: Department of Physics, Astrophysics, University of Oxford, Oxford OX1 3RH, UK\\
21: School of Physics and Astronomy, University of Southampton, Southampton SO17 1BJ, UK\\
22: Center for Astro, Particle and Planetary Physics, New York University Abu Dhabi, Abu Dhabi, United Arab Emirates\\
23: Department of Physics, Centennial Centre for Interdisciplinary Science, University of Alberta, Edmonton, AB T6G 2E1, Canada\\
24: East Asian Observatory, Hilo, Hawaii 96720, USA\\
25: Institute for Space Sciences, 077125 Bucharest-Magurele, Romania\\
26: Dr.~Karl Remeis-Sternwarte and Erlangen Centre for Astroparticle Physics, Friedrich-Alexander-Universit\"at Erlangen-N\"urnberg, 96049, Bamberg, Germany\\

\normalsize{$^\ast$To whom correspondence should be addressed; E-mail:  james.miller-jones@curtin.edu.au.}


\begin{sciabstract}
  The evolution of massive stars is influenced by the mass lost to stellar winds over their lifetimes.  These winds limit the masses of the stellar remnants (such as black holes) that the stars ultimately produce.  We use radio astrometry to refine the distance to the black hole X-ray binary Cygnus~X-1, which we find to be $2.22^{+0.18}_{-0.17}$ kiloparsecs. When combined with previous optical data, this implies a black hole mass of $21.2\pm2.2$ solar masses, higher than previous measurements.  The formation of such a high-mass black hole in a high-metallicity system constrains wind mass loss from massive stars. 
\end{sciabstract}


Gravitational wave detections of black hole merger events have revealed a population of black holes with masses ranging from 7 to 50 solar masses ($M_{\odot}$) \cite{Abbott2019}.  Black holes that interact with a companion star are visible to electromagnetic observations as an X-ray binary.  Radial velocity measurements of these companion stars have shown that black holes in X-ray binaries all have masses below $20M_{\odot}$ \cite{Casares2014}.  The highest measured black hole mass in an X-ray binary is $15.65\pm1.45 M_{\odot}$ for the extragalactic system M33 X-7  \cite{Orosz2007}.
  
The mass of a black hole is initially set by the properties of its progenitor star, augmented by accretion or mergers over its lifetime. The relevant properties of the progenitor include its initial mass and abundance of heavy elements (referred to as its metallicity), the mass lost in stellar winds over its lifetime, and the evolutionary pathway that it followed, which can be strongly influenced by a binary companion.   Mass measurements for massive stellar-mass black holes constrain stellar and binary evolution models \cite{Belczynski2010} and predictions of the expected black hole merger rates.

The X-ray binary Cygnus~X-1 (V1357 Cyg; coordinates in Table~S1) contains a black hole in a 5.6-day orbit with a more massive supergiant donor star, of spectral type O.  Previous estimates of its component masses were based on a parallax measurement, whereby the apparent annual angular shift in the source position relative to more distant objects was measured using radio very long baseline interferometry. Via trigonometry, this gave a distance of $1.86^{+0.12}_{-0.11}$\,kpc \cite{Reid2011}.  When combined with optical radial velocity measurements of the system, these data yielded a black hole mass of $14.8\pm1.0 M_{\odot}$ \cite{Orosz2011}. However, the derived system parameters are then inconsistent with the expected mass-luminosity relation for the donor, if it is a hydrogen-burning main sequence star \cite{Ziolkowski2014}.  The optical parallax measurement of $0.42\pm0.03$\,milliarcseconds (mas) using the {\it Gaia} space telescope \cite{Gaia2018}, after correction for the known zero-point offset of $\approx 0.05$\,mas (with estimates ranging from 0.03--0.08\,mas; \cite{Chan2020}), becomes $0.47\pm0.04$\,mas.  This is inconsistent with the radio value of $0.54\pm0.03$\,mas \cite{Reid2011}, and is unlikely to be due to orbital displacement of the donor star because the {\it Gaia} value is the average over 119 orbital periods.

Between May 29 and June 3, 2016, we performed six observations (one per day) of Cygnus~X-1 with the Very Long Baseline Array (VLBA) at 8.4\,GHz, sampling one full orbital period. To reduce systematic uncertainties in our position measurements, we phase referenced the data to a nearby calibrator source $0.4^{\circ}$ from Cygnus~X-1 \cite{SuppMaterial}. These data resolve the orbital motion of the black hole as projected onto the downstream surface from which the jet emission can escape, known as the photosphere. We find the orbit is clockwise on the plane of the sky (Fig.~1), in agreement with previous observations \cite{Reid2011}.

Combining the orbital phase coverage of our VLBA data with the archival observations \cite{Reid2011}, we simultaneously fitted the entire data set (covering a 7.4-year baseline) with a full astrometric solution incorporating linear motion across the sky (proper motion), parallax and orbital motion.  A fit to both right ascension and declination co-ordinates showed an orbital phase dependence in the direction of the residuals.  We attribute this to the effect of free-free absorption in the stellar wind, which is known to modulate the radio emission of Cygnus~X-1 on the orbital period \cite{Brocksopp2002}.  Electrons in the ionised wind of the O star can absorb radio photons in the presence of atomic nuclei, preventing the radio emission from the inner parts of the jet from reaching us.  This free-free absorption is reduced as the stellar wind density decreases on moving away from the O star, allowing radiation to escape from further downstream.  As the black hole (from which the jet is launched) moves around its orbit, the varying path length through the stellar wind imprints an orbital periodicity on the apparent radio position along the jet axis.  When the black hole is on the far side of the donor star, the path length and hence absorption are maximised, pushing the radio photosphere downstream along the jet axis \cite{SuppMaterial}.

To negate the impact of the stellar wind absorption, we therefore repeated our astrometric model fitting in one dimension only, perpendicular to the known jet axis. This removed the orbital phase dependence of the fit residuals perpendicular to the jet axis (Fig.~2).  Our measured semi-major axis of the black hole orbit is $58\pm20$ microarcseconds ($\micro$as), and our revised parallax measurement is $0.46\pm0.04$\,mas, consistent with the optical value from {\it Gaia} after correction for the zeropoint.  After converting our measured parallax to a distance using an exponentially-decreasing space density prior \cite{Astraatmadja2016}, we find a distance of $2.22^{+0.18}_{-0.17}$ kiloparsecs (kpc).

This revised distance affects the system parameters derived from the optical modelling \cite{Orosz2011}. We reanalysed the optical light curve \cite{Brocksopp1999a} and radial velocity curves \cite{Gies2003}, adopting our revised distance and additional constraints on the effective temperature, surface gravity and helium abundance of the donor star \cite{Shimanskii2012}. We find substantially higher masses for both the black hole ($21.2\pm2.2 M_{\odot}$; Fig.~S8) and the donor star ($40.6^{+7.7}_{-7.1}M_{\odot}$). The black hole orbital semi-major axis derived from this reanalysis of the optical data is $0.160\pm0.013$ astronomical units (au) (Table~1), which equates to $73\pm8$\,$\micro$as at our best-fitting distance of $2.22^{+0.18}_{-0.17}$\,kpc.  This is consistent with the value derived directly from our VLBA astrometry (Fig.~2, Table~S3).

The higher donor mass and greater luminosity inferred from the larger distance (Table~1) bring the system into closer agreement with the mass-luminosity relationship for main-sequence hydrogen-burning stars of solar composition \cite{Ziolkowski2014,SuppMaterial}.  However, the measured surface composition shows that the helium-to-hydrogen ratio is enhanced by a factor of 2.6 relative to the solar composition \cite{Shimanskii2012}.  This would imply a slightly different mass-luminosity relationship if the surface abundance is indicative of the overall composition of the donor star, but remains broadly consistent with our revised values (Fig.~3).

The higher mass and distance could also affect the black hole spin determined from spectral fitting of the X-ray continuum \cite{Gou2011}.  We therefore reanalysed archival X-ray data using a continuum-fitting method, assuming the black hole spin axis is aligned with the orbital plane. We find the black hole dimensionless spin parameter $a_{\ast}>0.9985$, close to the maximum possible value of 1, although this could be affected by systematic uncertainties \cite{SuppMaterial}. Even if the true value is less extreme, it would still be very high, consistent with previous results derived from both the continuum and iron line fitting methods \cite{Gou2011,Duro2016}.  While the spin derived from our analysis would be reduced if the black hole spin axis were not aligned with the orbital plane, even a $15^{\circ}$ misalignment would still require a high spin, $a_{\ast}=0.9696$.

As an X-ray binary system with a high-mass donor star, the black hole in Cygnus~X-1 cannot have been spun up by accretion from its companion at the maximum theoretical rate (known as the Eddington limit; $\sim 2 \times 10^{-7}\ M_\odot\ \mathrm{yr}^{-1}$ for Cygnus X-1), as that cannot occur for longer than the lifetime of the donor ($\sim 4$\,Myr for our inferred mass \cite{Bressan2012}). The accretion time may be close to the age of the jet-inflated nebula surrounding the source (a few tens of kyr; \cite{Russell:2007}). 
The current spin must therefore reflect the angular momentum of the core of the progenitor star.  An evolutionary pathway that could explain this is main sequence mass transfer from the black hole progenitor to the secondary star (Case A mass transfer; \cite{Kippenhahn1967}) with the core of the progenitor tidally locked and hence rapidly rotating, which can produce high black hole spins \cite{Qin2019}. This evolutionary pathway for Cygnus~X-1 would imply a spin axis of the black hole progenitor that is aligned with the orbital angular momentum.  Given the very low velocity kick of $9\pm2$\,km\,s$^{-1}$ imparted to the system on black hole formation (as determined by VLBA astrometry; \cite{Mirabel2003}), the black hole spin should still be aligned with the orbital inclination, as we assumed above. This scenario is also consistent with the low orbital eccentricity \cite{Orosz2011}, and the lack of strong quasi-periodic variability seen in the X-ray power density spectra of the source \cite{Grinberg2014}.

This evolutionary pathway is associated with enhanced nitrogen abundances, as observed in the donor spectrum \cite{Qin2019}.  The transfer of enriched material from the black hole progenitor could also explain the high helium abundance in the spectrum of the donor star \cite{Shimanskii2012}.  In the absence of convection, which is expected to be limited to a very thin surface layer in the envelope of a $40M_{\odot}$ star of solar (or possibly super-solar; \cite{Shimanskii2012}) metallicity, some fraction of this material would be retained on the surface for up to $\sim 10^5$\,yr (see Supplementary Text).  As a surface contaminant, this would not be reflective of the overall composition of the donor. Our revised values for the mass and luminosity of the donor star are consistent with theoretical expectations (Fig.~3, \cite{SuppMaterial}). 

The increase in the inferred black hole mass makes Cygnus~X-1 more massive than previously observed black holes in X-ray binaries \cite{Tetarenko2016}, surpassing M33 X-7.  M33 X-7 has a substantially sub-solar metallicity ($\sim0.1$ times the solar metallicity, $Z_{\odot}$; \cite{Orosz2007}). However, the metallicity of the mass donor in Cygnus~X-1 is much higher. It has been estimated to be approximately twice solar \cite{Shimanskii2012}, although the complexity of the system makes precise measurements challenging, and the true value may be closer to solar \cite{SuppMaterial}.
The existence of a $21M_{\odot}$ black hole at solar (or super-solar) metallicity implies that mass loss rate prescriptions \cite{Belczynski2010} over-estimate the mass loss during the luminous blue variable or Wolf-Rayet stages of stellar evolution \cite{Hurley2000,Vink2001}.
Assuming solar metallicity for the system, we find that either the mass loss rates in Wolf-Rayet winds from naked helium stars are reduced by a factor of three compared to current models \cite{Belczynski2010}, or those in luminous blue variable winds are reduced by at least a third, or both (see Supplementary Text).

Reduced mass loss can lead to higher progenitor masses at the time of the supernova.  It may also allow massive stars at moderate metallicities to retain hydrogen as they evolve (though probably not in the evolutionary history of Cygnus~X-1, where the black hole's progenitor likely had most of its hydrogen stripped off by the companion while the former was still on the main sequence; see Supplementary Text).  This would change the observational signatures of supernovae, for example producing hydrogen-rich (pulsational) pair-instability supernovae \cite{Arcavi:2017}.  Enrichment of the interstellar medium by stellar winds could be reduced, and, depending on which phases of the stellar evolution are most impacted by reduced winds, the contribution of massive stars to the re-ionization of the Universe may be affected (see Supplementary Text).

The black hole mass distribution inferred from gravitational wave events favours larger masses than predicted by stellar and binary evolution models.  Variations in metallicity-specific star formation history that favour greater star formation at lower metallicity have been proposed as an explanation \cite{Neijssel2019,Chruslinska2019}.  Reduced stellar winds would increase the mass of black holes that could be produced at all metallicities, such that massive gravitational wave sources could form at intermediate (not just very low) metallicities. This would imply that the progenitors of some gravitational wave events could have formed at correspondingly lower redshift, with a shorter delay time between binary formation and merger.

The high spin of Cygnus~X-1 (in common with most high-mass black hole X-ray binaries, which appear to be predominantly rapidly spinning; \cite{MillerMiller:2015}) implies that it followed a different evolutionary pathway to the majority of black holes detected in gravitational wave events, which have spins that are either low or misaligned \cite{Farr2017}.  Given the current orbital separation, we do not expect Cygnus~X-1 to undergo a binary black hole merger in a timescale comparable to the age of the Universe.



\section*{Acknowledgments}
{\bf Acknowledgments:} We acknowledge Guy Pooley's contribution to the radio observing campaign.  The Very Long Baseline Array is a facility of the National Science Foundation operated under cooperative agreement by Associated Universities, Inc.  This work made use of the Swinburne University of Technology software correlator, developed as part of the Australian Major National Research Facilities Programme and operated under licence.  This work has made use of data from the European Space Agency (ESA) mission
{\it Gaia} (\url{https://www.cosmos.esa.int/gaia}), processed by the {\it Gaia}
Data Processing and Analysis Consortium (DPAC,
\url{https://www.cosmos.esa.int/web/gaia/dpac/consortium}). Funding for the DPAC
has been provided by national institutions, in particular the institutions
participating in the {\it Gaia} Multilateral Agreement. {\bf Funding:} JCAM-J and IM are recipients of Australian Research Council Future Fellowships (FT140101082 and FT190100574, respectively) funded by the Australian government.  LJG acknowledges the support from the National Program on Key Research and Development Project through grant No. 2016YFA0400804, and from the National NSFC with grant No. U1838114, and by the Strategic Priority Research Program of the Chinese Academy of Sciences through grant No. XDB23040100.  VG is supported through the Margarete von Wrangell fellowship by the
ESF and the Ministry of Science, Research and the Arts Baden-W\"urttemberg.  BM acknowledges support from the Spanish Ministerio de Econom\'ia y Competitividad (MINECO) under grant AYA2016-76012-C3-1-P and from the Spanish Ministerio de Ciencia e Innovaci\'on under grants PID2019-105510GB-C31 and CEX2019-000918-M of ICCUB (Unidad de Excelencia ``Mar\'ia de Maeztu'' 2020--2023). SM was supported by the Netherlands Organization for Scientific Research (NWO) VICI grant (no. 639.043.513).  GRS acknowledges support from an NSERC Discovery Grant (RGPIN-2016-06569). VT is supported by programme Laplas VI of the Romanian National Authority for Scientific Research.  JW acknowledges funding from the Bundesministerium f\"ur Wirtschaft und Technologie under Deutsches Zentrum f\"ur Luft- und Raumfahrt grant 50\,OR\,1606.  JZ acknowledges the support from the Polish National Science Centre grant 2015/18/A/ST9/00746. {\bf Author contributions:} JCAM-J analyzed the VLBA data, and led the manuscript preparation.  AB conducted the astrometric fitting. JAO conducted the optical light curve and radial velocity curve fitting.  IM and CJN performed the stellar wind modelling and led the discussion of the system's evolution, with input from TJM. LG, XZha and XZhe performed the analysis of the black hole spin.  JZ calculated the mass-luminosity tracks.  MJRe provided the archival VLBA data and guidance on precision astrometry.  PU, VG, JCAM-J and JW co-ordinated the VLBA observing campaign, with theoretical input from SM.  
JCAM-J, PU, TJM, VT, APR, JW and DMR wrote the observing proposal, and 
D-YB, RD, TJ, J-SK, BM, MJRi, GRS, and AJT contributed to the design and setup of the observations. All authors provided input and comments on the manuscript. {\bf Competing interests:} The authors declare no conflicts of interest.  {\bf Data and materials availability:} The raw VLBA data are available from the NRAO archive  (\url{https://archive.nrao.edu/archive/advquery.jsp/}), under project codes BR141 and BM429.  Our measured positions are listed in Table S1. The COMPAS population synthesis code we used is available at \url{http://github.com/TeamCOMPAS/COMPAS}. The software for performing the astrometric and optical model fitting, the spin fitting, and for calculating the mass-luminosity relationships, together with our COMPAS input and output files, is available in our code repository at \url{https://github.com/bersavosh/CygX-1_JMJ2020}, which is archived at \url{https://zenodo.org/record/3961240} \cite{Miller-Jones2021}.

\clearpage

\renewcommand{\thefigure}{{\bf 1}}
\begin{figure}
\centering
\includegraphics[width=\textwidth]{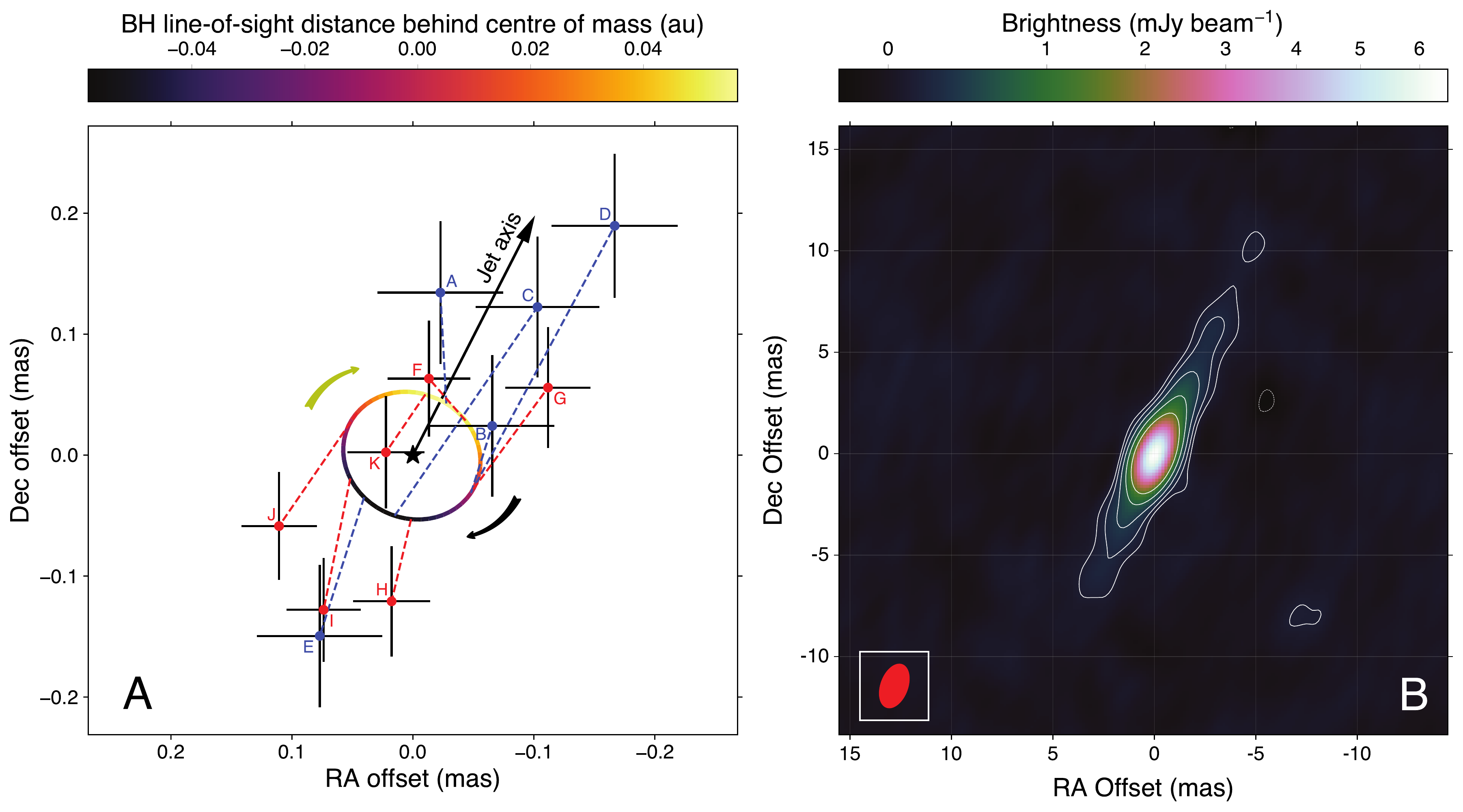}
\caption{{\bf Cygnus X-1 and its best fitting model orbit on the plane of the sky.}  (A) astrometric measurements from the VLBA (red points) and the archival data from \cite{Reid2011} (blue points). Error bars show the 68\% confidence level.  The letter labels reflect the chronological ordering of the observations, as detailed in Table~S1.  Dashed lines link the measured positions to the location on the fitted orbit, shown as the colored ellipse.  Color bar indicates the location of the black hole along the line of sight, relative to the centre-of-mass of the system (shown as the black star), with positive values being behind the centre of mass.  Arrows indicate the direction of orbital motion and the jet axis. (B) stacked radio image of the jet in color, with white contours every $\pm(\sqrt{2})$ times the rms noise level of 23 microJanskys ($\micro$Jy) per beam. Red ellipse shows the synthesised beam. Although the measured positions scatter along the jet axis, the motion perpendicular to the jet axis is reproduced by the astrometric model (see Fig.~2). Coordinates are given in right ascension (RA) and declination (Dec), J2000 equinox.\label{fig:fig1}}
\end{figure}

\renewcommand{\thefigure}{{\bf 2}}
\begin{figure}
\centering
\includegraphics[height=\textwidth,angle=270]{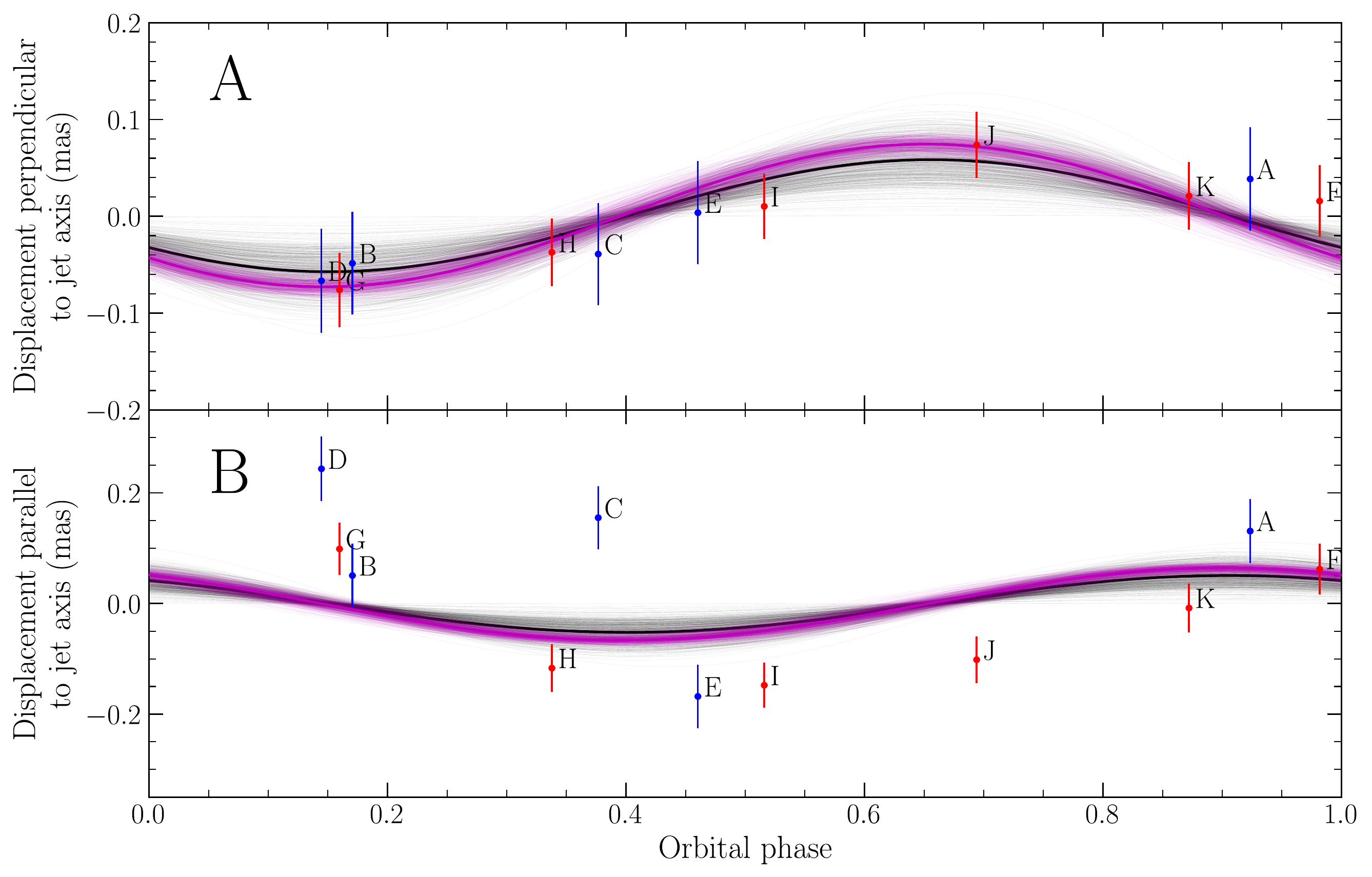}
\caption{{\bf Orbital displacements relative to the best-fitting one-dimensional astrometric model.} Parallax and proper motion signatures have been subtracted. (A) The measured displacements perpendicular to the jet axis. Red points are our VLBA data, and blue points are the archival observations \cite{Reid2011}, with error bars showing the 68\% confidence level. Labels reflect the chronological ordering of the observations, as listed in Table~S1. Black and magenta lines show 500 random draws from the posterior probability distribution of the orbital parameters, for radio and optical models, respectively (with the posterior medians indicated by thicker lines), which are consistent within the uncertainties. The data were only fitted perpendicular to the jet axis.  (B) The measured displacements parallel to the jet axis show that the measured core positions are primarily downstream of the model predictions when the black hole is close to superior conjunction (behind the donor star; phases close to 0.0), and upstream when the black hole is close to inferior conjunction (phases close to 0.5), as expected for wind absorption.\label{fig:fig2}}
\end{figure}

\renewcommand{\thefigure}{{\bf 3}}
\begin{figure}
\centering
\includegraphics[width=\textwidth]{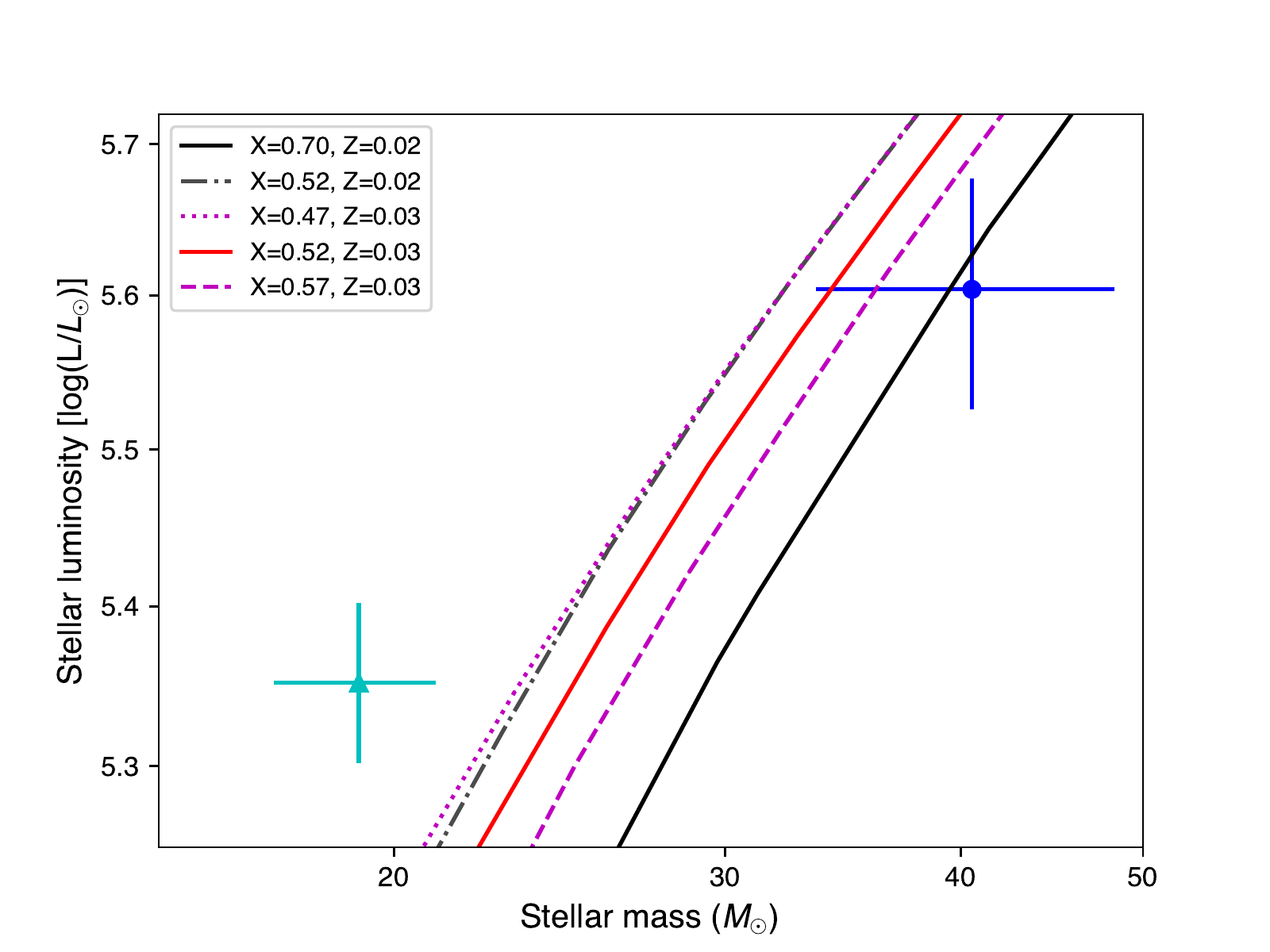}
\caption{{\bf Predicted mass-luminosity relations for high-mass main sequence stars.} Masses are given in solar masses, and luminosities relative to the solar luminosity $L_{\odot}$. The black solid line shows the predicted relation for a standard composition (hydrogen mass fraction $X=0.70$; mass fraction of heavy elements $Z=0.02$ \cite{SuppMaterial}). The grey dot-dashed line is for an enhanced helium abundance, $X=0.52$, $Z=0.02$ (as inferred for the surface abundance of the donor star; \protect{\cite{Shimanskii2012}}). The red solid line shows the effect of an increased metallicity, with $Z=0.03$, $X=0.52$. The magenta dotted and dashed lines show the effect of the uncertainty on the helium abundance \protect{\cite{Shimanskii2012}}. The mass and luminosity determined from previous observations \protect{\cite{Orosz2011}} are shown as the cyan triangle. The values from this work are shown as the blue circle, which lies closer to the theoretical relations, irrespective of composition or metallicity. Error bars show 68\% confidence levels.\label{fig:fig3}}
\end{figure}

\clearpage

\renewcommand{\thetable}{{\bf 1}}
\begin{table}
    \centering
 \begin{tabular}{l c c c c} 
 \hline
 Parameter & Median & Mode & Lower bound & Upper bound \\ [0.5ex] 
 \hline
$i$ (deg) & 27.51 & 27.33 & 26.94 & 28.28 \\
$e$ & 0.0189 & 0.0186 & 0.0163 & 0.0217 \\
$\omega$ (deg) & 306.6 & 306.3 & 300.3 & 313.1 \\
$M_{1}$ ($M_{\odot}$) & 40.6 & 39.8 & 33.5 & 48.3 \\
$f_{1}$ & 0.960 &  0.999 & 0.930 & 0.988 \\
$T_{\rm eff}$ (K) & 31,138 & 31,158 & 30,398 & 31,840 \\
$K_{1}$ (km s$^{-1}$) & 75.21 & 75.18 & 74.80 & 75.63 \\
$\phi$ & 0.0024 & 0.0023 & 0.0013 & 0.0034 \\
$\Omega_{\rm rot}$ & 1.05 & 1.04 & 0.95 & 1.16 \\
\hline
$M_{\rm BH}$ ($M_{\odot}$) & 21.2 & 21.4 & 18.9 & 23.4 \\
$R_{1}$ ($R_{\odot}$) & 22.3 & 22.2 & 20.6 & 24.1 \\
$\log L/L_{\odot}$ & 5.625 & 5.606 & 5.547 & 5.698 \\
$\log (g_{1} / {\rm cm\,s}^{-2})$ & 3.348 & 3.351 & 3.335 & 3.360 \\
$a$ (au) & 0.244 & 0.243 & 0.231 & 0.256 \\
$a_{1}$ (au) & 0.0838 & 0.0840 & 0.0816 & 0.0856 \\
$a_{\rm BH}$ (au) & 0.160 & 0.159 & 0.147 & 0.173 \\
\hline
\end{tabular}
\caption{{\bf Fitted and derived physical parameters for Cygnus~X-1.} The inclination $i$, eccentricity $e$, and argument of periastron $\omega$ of the orbit; the mass $M_{1}$, Roche-lobe filling factor $f_{1}$, effective temperature $T_{\rm eff}$, and semi-amplitude of the radial velocity curve $K_{1}$ for the O-star; a phase shift parameter $\phi$ to account for ephemeris errors; and the ratio $\Omega_{\rm rot}$ of the rotational frequency of the O-star to the orbital frequency (the nine parameters above the horizontal line) were directly fitted in the model.  The black hole mass $M_{\rm BH}$; the radius $R_{1}$ in solar radii $R_{\odot}$, luminosity $L$ in solar luminosities $L_{\odot}$,
and surface gravity $g_{1}$ of the O-star; and the semi-major axes of the full orbit $a$, stellar orbit $a_{1}$, and black hole orbit $a_{\rm BH}$ (the seven parameters below the horizontal line) were derived from the fitted parameters, using the known orbital period \protect{\cite{Brocksopp1999a}}. Lower and upper bounds encompass the 68\% confidence interval.}
\label{table:elc_fitparms}
\end{table}

\clearpage

\section*{Supplementary Materials}

{\bf This PDF file includes}\\
\noindent
Materials and Methods\\
Supplementary Text\\
Figures S1 to S11\\
Tables S1 to S3\\
References \textit{(33-118)}

\clearpage

\section{Materials and Methods}

Our radio observations were made using the VLBA, as part of the Cygnus~\X-1 Hard-state Observations of a Complete Binary Orbit in X-rays (CHOCBOX) project; a multi-wavelength observing campaign to monitor Cygnus~\X-1 around an entire binary orbit.

\subsection{Calibrator search}

The closest two known compact extragalactic radio calibrator sources \cite{Ma2009} to Cygnus~\X-1 are NVSS~J195330+353759 (hereafter J1953+3537) and NVSS~J195740+333827 (hereafter J1957+3338), which lie 1.08\degr\ and 1.57\degr\ away, respectively (see Fig.~S1). We chose J1957+3338 as our primary phase reference calibrator. Despite being further from Cygnus~\X-1, it is less resolved at our observing frequency of 8.4\,GHz, allowing better calibration of the longer baselines to stations at Mauna Kea and St Croix, and hence better astrometric precision.

Because systematic errors in astrometry scale with the distance between the target and the phase reference source \cite{Pradel2006}, we conducted a calibrator search to find a closer phase reference source and thereby reduce the systematic uncertainties.  We observed with the VLBA on 2016 April 9, targeting 44 sources from the National Radio Astronomy Observatory (NRAO) Very Large Array (VLA) Sky Survey (NVSS) catalog \cite{Condon1998} within $30^{\prime}$ of Cygnus~\X-1.  Observing at 1.64\,GHz, the primary beam was large enough to cover the 44 targets in 4 pointings.  We used the multi-phase centre capability of the DiFX software correlator \cite{Deller2011} to provide individual data sets for each target.  Our observing bandwidth was 64\,MHz, and we achieved roughly 8\,min of observations on each target source.  Due to the scattering along the line of sight to the Cygnus region we did not use the antennas at Mauna Kea or St Croix.  The brightest source detected (NVSS~J195754+353513, hereafter J1957+3535), at $7.7\sigma$, was located at co-ordinates (J2000) 19$^{\rm h}$57$^{\rm m}$54\fs105, $+$35$^{\rm d}$35$^{\prime}$13\farcs02, $0.4^{\circ}$ from Cygnus~\X-1.  It appeared unresolved to our observations, with a peak flux density of 2.8\,mJy\,beam$^{-1}$.  We therefore selected this source as a secondary phase reference source for our observing campaign.

\subsection{VLBA observations}

We observed Cygnus~\X-1 daily from 2016 May 29 through June 3, using the VLBA under project code BM429.  Our observations covered an entire 5.6-day binary orbit, allowing us to track the orbital motion on the plane of the sky.  Each observation was taken at a central frequency of 8.416\,GHz, with a bandwidth of 256\,MHz.  We cycled between Cygnus~\X-1, our primary phase reference calibrator J1957+3338 and our secondary phase reference source J1957+3535, spending 100, 60 and 60\,s per cycle, respectively, on each source.

Our observing runs were 12\,hr in duration, with half-hour geodetic blocks at the beginning, middle and end of each observation.  In each of these blocks, we observed a range of bright calibrator sources spread across the sky to solve for unmodeled tropospheric delay and clock errors, to improve our astrometric precision.

Our data were correlated using the DiFX software correlator \cite{Deller2011}, and reduced according to standard procedures within the 31DEC17 version of the Astronomical Image Processing System (\textsc{aips}; \cite{Greisen2003}).  Following initial fringe fitting on J1957+3338, we stacked the data from all six epochs to provide a global model that was used for the final fringe fitting.  The final phase, delay and rate solutions were interpolated to both the target source and the secondary phase reference calibrator, J1957+3535.  To improve our amplitude calibration, we also self-calibrated the data on J1957+3338 from each epoch in amplitude and phase, on a 10-minute timescale, using our global model.  We interpolated the resulting amplitude self-calibration solutions only (having zeroed the phases to prevent them from overriding the shorter-timescale phase solutions derived from the fringe fitting) to both the target source and J1957+3535.  We then imaged both Cygnus~\X-1 and J1957+3535 at each epoch, and measured their positions by fitting the core regions with elliptical Gaussians in the image plane. Stacked images of the two sources made with all six epochs of data are shown in Figs.~1 and S2, respectively.

\subsection{Archival data}

Because our own data were taken over a period of just six days, they could not constrain the parallax and proper motion of Cygnus~\X-1.  However, they fully sample the 5.6-day orbit, so when combined with archival data from 2009--2010 \cite{Reid2011} (VLBA program code BR141), can measure the parallax and refine the proper motion via the extended time baseline.  These archival data also provide an additional five samples of the binary orbit.

\subsection{Systematic uncertainties}
\label{sec:systematics}

With a signal-to-noise $>100$, our astrometric measurements were limited by systematic rather than statistical uncertainties.  These include contributions from the ionosphere \cite{Reid2017} and the troposphere \cite{Pradel2006}, both of which scale with angular separation between target and phase calibrator source.

Ionospheric wedges can produce parallax gradients across the sky, leading to systematic shifts of order 50\,\micro as\,($\nu/6.7$\,GHz)$^{-2}$ per degree of offset between target and phase reference source, where $\nu$ is the observing frequency \cite{Reid2017}.  While this effect can in principle be minimized by fitting an ionospheric wedge using multiple phase reference sources surrounding the target, none of our Cygnus~\X-1 data sets contained sufficient well-situated calibrators to enable such an approach, so we adopted the above estimate of the likely systematic uncertainty.

For our VLBA data, we used the measured offsets of the nearby secondary phase reference calibrator, J1957+3535, from its mean position to correct the astrometry, combining the statistical uncertainties from each source in quadrature.  This effectively phase referenced Cygnus~\X-1 to J1957+3535, with the 0.4$^{\circ}$ calibrator offset implying expected systematic uncertainties from the troposphere of 23 and 29\,\micro as in RA and Dec., respectively \cite{Pradel2006}.  We flagged local sunrise at each station to prevent ionospheric wedges from affecting the astrometric measurements, which, together with the long (12-hour) tracks, reduces the effects of any ionospheric wedge.  With a calibrator offset of 0.4$^{\circ}$, the estimated ionospheric systematic uncertainty was 13\,\micro as, negligible compared to the tropospheric and combined statistical uncertainties.

The archival observations \cite{Reid2011} used two different phase calibrator sources, J1953+3537 and J1957+3338, located $1.08^{\circ}$ and $1.57^{\circ}$ from Cygnus~\X-1, respectively (see Fig.~S1).  We used their position measurements \cite{Reid2011}, taking a mean of the two measurements (i.e.\ the positions measured relative to the two different phase reference sources) at each epoch to minimise the effect of any ionospheric wedge, and adding the statistical uncertainties from the two different measurements in quadrature.  To this we added in quadrature both the estimated tropospheric systematics (38 and 47\,\micro as in RA and Dec., respectively; \cite{Pradel2006}), and the estimated ionospheric systematic uncertainties of 34\,\micro as in both dimensions.  While the dependence of ionospheric systematic uncertainties on observing duration and time of year is not yet understood, we regard these estimates as conservative. The final positions of Cygnus~X-1 and their uncertainties are given in Table~S1. 

The above discussion outlines the systematic uncertainties affecting the relative astrometry measurements on which our results are based.  To enable comparisons with other work, we also estimate the uncertainties affecting our absolute astrometric positions. J1957+3338 is one of the sources making up the third realization of the International Celestial Reference Frame (ICRF3).  However, all our positional measurements were derived assuming the position of J1957+3338 to be 19$^{\rm h}$57$^{\rm m}$40\fs 5499147, 33\degr38$^{\prime}$27\farcs 943527, which is offset by 96\micro as in RA and 323\micro as in Dec from the ICRF3 positions \cite{Charlot2020}.  The uncertainties on the ICRF3 positions are 151\micro as in RA and 181\micro as in Dec.  The uncertainties on the absolute positions of Cygnus X-1 (Tables~S1 and S3) and J1957+3535 \cite{SuppMaterial} will be further augmented by the systematic uncertainties due to the offset between phase reference calibrator and target positions, as discussed above.  Adding these contributions in quadrature, our absolute astrometric positions are uncertain by 181\micro as in RA and 375\micro as in Dec.

\subsection{Astrometric model fitting}
\label{sec:standard_fit}
The orbital motion of the black hole can be measured in astrometric VLBI observations \cite{Reid2011}. Those observations showed that the sense of the black hole orbit was clockwise on the sky, but only measured the size of the orbit to $2\sigma$, as $0.18\pm0.09$\,au.  Our VLBA data, corrected for the nominal parallax and proper motion over the six days of the campaign, showed a clockwise orbit that was resolved on the plane of the sky (Fig.~1).  Our measured source positions are those of the radio core of the jet, which is presumably launched from close to the black hole.  However, the radio photosphere is located $\approx$ 0.5--5\,mas downstream of the black hole itself \cite{Szostek2007,Tetarenko2019}, so we see the projection of the true black hole orbit on the radio photosphere, potentially modified by orbitally- or wind-induced bending of the jet axis \cite{Szostek2007,Yoon2015}.

To sample both the orbit and proper motion of the system, we combined our VLBA data with the archival observations \cite{Reid2011}, and fitted the combined data set with a model incorporating proper motion, parallax and orbital motion.  A full astrometric solution includes the reference position in both co-ordinates ($\alpha_0$ and $\delta_0$), the proper motion in both co-ordinates ($\mu_{\alpha}\cos\delta$ and $\mu_{\delta}$), the parallax ($\pi$), plus the seven orbital parameters (orbital period $P$, epoch of periastron passage $T_0$, eccentricity $e$, inclination $i$, semimajor axis $a_{\rm BH}$, argument of periastron $\omega$, and longitude of the ascending node, $\Omega$).  We adopted the orbital period $P$ from the photometric ephemeris \cite{Brocksopp1999a}.  
Constraints on the inclination of the binary orbit, the eccentricity, the argument of periastron, and the masses of the two components were taken from previous work \cite{Orosz2011}, which, with the orbital period, can be converted to a constraint on the semimajor axis via Kepler's Third Law. For the known inclination angle and argument of periastron \cite{Orosz2011}, periastron passage (which defines orbital phase zero in their formalism) occurs at phase 0.109 in an alternative ephemeris \cite{Brocksopp1999a}, which assumed a circular orbit and defined orbital phase zero to occur at superior conjunction of the black hole \cite{Zanin2016}.  We therefore defined our reference epoch $T_0$ to be the reference date for superior conjunction determined by \cite{Brocksopp1999a}, augmented by this phase shift.

To fit the motion of the system on the sky, we used a Markov chain Monte Carlo (MCMC) approach, using the \textsc{pymc3} package \cite{Salvatier2016}.  We adopted the Hamiltonian Monte Carlo formalism (HMC; \cite{Neal2012,Betancourt2017}) with a No-U-Turn Sampler (NUTS; \cite{Hoffman2011}).  We used Gaussian priors on $i$ and $\omega$ based on the optical modelling work of \cite{Orosz2011}. Because optical modelling cannot determine the sense of the orbit as seen on the plane of the sky, optically-determined inclinations of $i$ and $(180^{\circ}-i)$ are degenerate, with values of $90^{\circ}<i<180^{\circ}$ corresponding to the clockwise orbit seen in our VLBA data. Furthermore, the two components of the binary each have their own argument of periastron, $\omega_{\ast}$ and $\omega_{\rm BH}$, which are separated by $\omega_{\ast} = \omega_{\rm BH}+180^{\circ}$.  Using the constraints from \cite{Orosz2011}, we therefore adopted a prior of $152.94\pm0.76^{\circ}$ for the orbital inclination, $127.6\pm5.3^{\circ}$ for the argument of periastron of the black hole $\omega_{\rm BH}$, and uniform priors on six of the other seven parameters ($\alpha_0$, $\delta_0$, $\mu_{\alpha}\cos\delta$, $\mu_{\delta}$, $\pi$, and $a_{\rm BH}$).  For the final parameter, $\Omega$, we assumed that the jet was aligned with the orbital angular momentum vector.  The jet position angle is known on the plane of the sky, so with this constraint and the inclination and argument of periastron from the optical modelling work, we infer the longitude of the ascending node to be $64^{\circ}$.  We therefore used a Gaussian prior on $\Omega$ with this mean, and a standard deviation of $1^{\circ}$.  A summary of our adopted priors is given in Table~S2.

This astrometric model fitting gave a parallax of $\pi = 535\pm28$\,\micro as and an orbital semi-major axis for the black hole of $a_{\rm BH} = 89\pm15$\,\micro as (see Table~S3 for full results). The fit residuals are shown in Fig.~S3, and are substantially larger than we expected from our (conservative) estimates of the uncertainties on the data.

\subsection{Wind absorption}
\label{sec:1d_fit}
To achieve a reduced $\chi^2$ value of 1 for their astrometric model fitting, \cite{Reid2011} found that they had to adopt error floors of 80 and 160 microarcseconds (\micro as) in RA and Dec., respectively, which is again substantially larger than expected from the calibrator throw.  This suggests an extra source of systematic uncertainty in addition to those modelled by simulations \cite{Pradel2006}.

Cygnus~\X-1 has a strong stellar wind, which attenuates the radio emission at frequencies above 2\,GHz \cite{Brocksopp2002}.  The strong free-free absorption by the wind is enhanced at superior conjunction of the black hole, which is then seen through the maximum path length through the wind.  At inferior conjunction, the path length is minimized and the radio emission is least absorbed.  Because we are observing at a frequency of 8.4\,GHz, where the orbital modulation of the radio emission is $13.8\pm2.4$\% \cite{Szostek2007}, our data are affected by wind absorption.  Close to superior conjunction of the black hole, this would have the effect of pushing the optical depth $\tau=1$ surface out to larger distances downstream.  Changes in the location of the $\tau=1$ surface due to intrinsic variations in the electron number density or magnetic field strength in the jet would lead to additional, non-orbital shifts in the measured position of the radio core.  

The jet axis in Cygnus~\X-1 is well-known, oriented at $-26^{\circ}$ east of north (Fig.~1), and stable over time \cite{Stirling2001,Reid2011}.  We can therefore determine the positions measured parallel and perpendicular to the jet axis, for each epoch.  Because the wind absorption and intrinsic jet variability should only affect the parallel co-ordinate, we can conduct astrometric fits only on the perpendicular component to provide an estimate of the true astrometric and orbital parameters of the system \cite{Miller-Jones2021}.  However, when fitting in only one dimension, the reference position ($\alpha_0$, $\delta_0$) and proper motion ($\mu_{\alpha}\cos\delta$, $\mu_{\delta}$) become degenerate in the two sky co-ordinates.  To overcome this problem, we used Gaussian priors for both $\alpha_0$ and $\mu_{\alpha}\cos\delta$, with the means and standard deviations taken from the posterior probability distributions of the original model fitting.

Using this model gives a substantially smaller parallax of $\pi=458\pm35$\,\micro as (Fig.~S4), and a semi-major axis for the black hole orbit of $a_{\rm BH}=58\pm20$\,\micro as.  Our final best-fitting parameters are given in Table~S3.  Fig.~S5 shows the one-dimensional histograms of the posterior probability distribution of each of the nine fitted parameters, together with two-dimensional scatter plots showing the covariances of the parameters. The correlations between parameters arise due to the use of the one-dimensional fit.  There is a degeneracy in position along the jet axis, as reflected in the mean value of the residuals parallel to the jet axis, which we found to be $86$\,\micro as upstream (i.e.\ towards the south south-east on the sky).  This is well within the peak of the two-dimensional posterior probability distributions of ($\alpha_0,\delta_0$) shown in Fig.~S5.  This shift has been corrected in Fig.~1, and also in Fig.~2, which shows the best-fitting displacements both perpendicular and parallel to the jet axis.

The reduction in the amount of information available (fitting in one co-ordinate rather than two) increases the uncertainties in the fitted parameters.  However, this model is based on well-characterised properties of the system, and reduces the systematic uncertainty caused by the radio photosphere moving up and down the jet axis as a function of orbital phase.  Figs.~1 and 2 show that, as we expected, the measured position is scattered further downstream along the jet axis at superior conjunction of the black hole, and back towards the black hole at inferior conjunction.

Our measured parallax of $458\pm35$\,\micro as is consistent with the zero-point corrected optical value of $\sim470\pm40$\,\micro as from the second data release (DR2) from the {\it Gaia} space telescope\cite{Gaia2018,Chan2020}, which was recently refined to $468\pm15$\,\micro as in the early version of the third data release (eDR3; \cite{Gaia2020,Lindegren2020}). Unless the uncertainty is very small, simple inversion of a measured parallax to determine a source distance can introduce a non-negligible bias in the distance estimation.  With a 7.7\% parallax uncertainty, we therefore adopted the standard Bayesian formalism \cite{Luri2018} to convert our measured parallax to a probability density function for the source distance, using an exponentially-decreasing space density prior \cite{Astraatmadja2016}. We find a median distance of 2.22\,kpc, with a $1\sigma$ range of 2.05--2.40\,kpc, and a 90\% Bayesian credible interval of 1.96--2.54\,kpc.

\subsection{Spin-orbit misalignment}
\label{sec:alignment}

The proper motion of Cygnus~\X-1 relative to its likely parent stellar association Cygnus OB3 is $10.7\pm2.7$\,km\,s$^{-1}$ \cite{Rao2020}, indicating that it formed with little to no natal kick, and possibly without a supernova explosion \cite{Mirabel2003}. Our revised distance does not change this conclusion, because it still falls within the distance range of $2.0\pm0.3$\,kpc determined for Cyg OB3 \cite{Rao2020}.

In the absence of an asymmetric supernova kick, we would expect the angular momentum vector of a rapidly-spinning black hole like Cygnus~\X-1 \cite{Gou2011,Gou2014}, and hence the jet axis, to be aligned with the orbital angular momentum vector, which motivates our choice of a highly-constrained Gaussian prior for the longitude of the ascending node, $\Omega$. From the observed drift velocity of Cygnus~\X-1 relative to Cygnus OB3, we estimate the maximum asymmetric natal kick that the system could have received as 10--20\,km\,s$^{-1}$.  The maximum likely misalignment angle should scale as the ratio of kick velocity to the pre-supernova orbital velocity.  For the observed mass ratio between black hole and donor star, and assuming a pre-supernova orbital velocity of $\sim300$\,km\,s$^{-1}$, then unless the asymmetric kick was directly opposed to the Blaauw kick from mass ejection \cite{Blaauw1969}, we estimate a maximum misalignment angle of $\sim10^{\circ}$.  While inner disk misalignments of this size have been inferred from some X-ray spectral fitting results \cite{Tomsick2014,Parker2015,Walton2016}, other work suggests that this conclusion is sensitive to the assumed electron density in the accretion disk, with no misalignment being required in the case of the higher densities that would be appropriate for X-ray binary disks \cite{Tomsick2018}. 

To quantify the effect of any misalignment, we explored the effect of increasing the standard deviation on the Gaussian prior on $\Omega$ to $15^{\circ}$.  While the the posterior distribution for $\Omega$ increased to $77\pm12^{\circ}$, all other changes were well within the uncertainties.  The parallax increased by less than one percent, to $461\pm35$\,\micro as.  The semi-major axis of the black hole orbit increased to $63\pm20$\,\micro as, the argument of periastron increased by a degree to $126\pm5^{\circ}$, and the inclination did not change.  Even for the most extreme case of a uniform prior on $\Omega$ (0--360\degr), when the posterior distribution on $\Omega$ increased to $95\pm18^{\circ}$, there was minimal further change in the parallax, with a median and 68\% confidence interval of $464\pm35$\,\micro as.  Even then, the case of alignment still falls within the 90\% Bayesian confidence interval.  We conclude that our distance determination is therefore insensitive to the prior on $\Omega$, and that the data are consistent with an aligned jet axis.

\subsection{Updated Dynamical Model}\label{lcfit}

The increase in the distance of Cygnus~\X-1 compared to the previous results \cite{Reid2011} implies that the absolute magnitude of the donor star is larger, in turn increasing the inferred stellar radius.  The radius is an input to the dynamical model \cite{Orosz2011}, which used optical radial velocity and light curves, a measurement of the projected rotational velocity of the O-star and a measurement of the radius of the O-star, to determine the physical parameters of the system.  Given the revised stellar radius arising from our distance determination, we therefore reanalysed the optical photometric and velocity curves of \cite{Brocksopp1999b} and \cite{Gies2003} to update the estimated system parameters.

We retained the previous stellar rotational velocity estimate (taken from \cite{Caballero-Nieves2009}).  The radius and luminosity of the O-star can be computed as a function of temperature \cite{Orosz2011}, as shown in Fig.~S6, where for simplicity
we have adopted the revised distance with a conservative and symmetric uncertainty, $2.22\pm 0.18$\,kpc.  The larger distance implies larger radii and higher luminosities as compared to the previous results \cite{Orosz2011}.  We also used optical spectra \cite{Shimanskii2012} to improve previous estimates of the effective temperature, surface gravity and helium abundance of the O-star, which further constrain the stellar radius and luminosity.

To model the light curve and radial velocity curve we used the {\sc elc} code \cite{Orosz2000}, choosing the optimizer code based on the differential evolution (DE-MCMC)
algorithm \cite{TerBraak2006}.  We adopted the
``Model D'' framework from
\cite{Orosz2011}, which incorporates nonsynchronous rotation of the O-star
and an eccentric orbit.  We modified Model D \cite{Orosz2011} to fit for the Roche lobe filling factor rather than the radius of the O-star.  We define this filling factor as the ratio of the distance between the centre of the star and the point on its surface closest to the black hole, to the distance between the centre of the star and the L1 Lagrange point.  Our model therefore has 9 free parameters $(i,K_1,M_{1},f_1,\phi,e,\omega,\Omega_{\rm rot},T_{\rm eff})$.  We used the same 
table of model atmosphere specific intensities as \cite{Orosz2011}, and hence no parameterized limb darkening law is needed.

Because the donor star is close to Roche lobe filling \cite{Orosz2011}, the Roche geometry places strong constraints on the surface gravity, which is close to $\log g_1=3.3$ \cite{Orosz2011}. Stellar surface gravity and temperature are partly degenerate in optical spectral fitting \cite{Caballero-Nieves2009}, so this places constraints on the likely effective temperature. While there are several estimates of the effective temperature of the donor star in the literature \cite{Shimanskii2012,Herrero1995,Caballero-Nieves2009,Ziolkowski2005}, we discarded \cite{Herrero1995}, because it did not include the effects of line blanketing, which has a substantial impact on the effective temperature.  Of the remainder (i.e., \cite{Shimanskii2012,Ziolkowski2005,Caballero-Nieves2009}), all suggested the effective temperature to be in the range 30,000--31,000\,K (where we used the green spectrum of \cite{Caballero-Nieves2009}, as their blue spectrum gave an inconsistent estimate of the stellar surface gravity).  
We consider a broad weighted mean of these three effective temperature determinations, giving $T_{\rm eff}=30,200\pm900$\,K.

We used uniform priors for the 9 fitting parameters, including a broad prior on $T_{\rm eff}$ of 27,500--36,000\,K. Our likelihood function is based on the $\chi^2$ statistic and has two parts
\begin{equation}
\chi^2_{\rm total} = 
\chi^2_{\rm data} +
\chi^2_{\rm constraints}.
\end{equation}
$\chi^2_{\rm data}$ applies to the $U$, $B$, and
$V$ light curves and the radial velocity curve, [\cite{Orosz2011}, their equation (2)].  Five additional measured properties of Cygnus~\X-1 are independent of orbital phase, namely the radius of the O-star, $R_1(T_{\rm eff})$, which is computed from the parallax for a given temperature; the rotational velocity of $V_{\rm rot}\sin i=96\pm 6$\,km s$^{-1}$ \cite{Caballero-Nieves2009}; our weighted mean for the O-star temperature of $30,200\pm 900$ K; the O-star surface gravity of $\log g_1=3.31\pm 0.05$ \cite{Shimanskii2012}; and the lack of an X-ray eclipse.  A given vector of the 9 model parameters will produce values for all of these observed parameters, and the second part of the likelihood function is then found by
\begin{align}
\chi^2_{\rm constraints}&=
\left(\frac{{R_1-R_1(T_{\rm eff})}}{\sigma_{R_1(T_{\rm eff})}}   \right)^2
+
\left(\frac{{V_{\rm rot}\sin i-96}}{6}   \right)^2\nonumber\\
&+
\left(\frac{{T_{\rm eff}-30,200}}{900}   \right)^2
+
\left(\frac{{\log g -3.31}}{0.05}   \right)^2
+
\Theta_{\chi^2},
\end{align}
where $\Theta_{\chi^2}=10^6$ if there is an X-ray
eclipse, and zero otherwise.  The uncertainty
of $R_1(T_{\rm eff})$ depends on the
effective temperature, hence we used $\sigma_{R_1(T_{\rm eff})}$
in the expression above.  The inclusion
of $T_{\rm eff}$ in the expression for
$\chi^2_{\rm constraints}$ effectively imposes a Gaussian
prior for that parameter.

\subsection{{\sc elc} Model Fitting}
\label{sec:elc}

To find the uncertainties on the fitted and derived parameters,  we implemented the DE-MCMC algorithm \cite{Miller-Jones2021}. The code was
run 8 times using 40 chains each.
The chains were initialized using slightly tweaked copies
of the nearly optimal model (based on the results of \cite{Orosz2011}), and each run had a different
initial seed in the random number generator.  
The individual runs were done on computers with different
CPU speeds, with between 2,500 and 26,000 generations for each run.

To determine the burn-in time (how long the chains need to disperse from the initial state), we visually inspected plots of the chains over the first few hundred generations, finding that it took between about 20 and 30 generations for the chains to initially disperse. Thereafter, the chains reached a steady-state by generation 150 or so. To be conservative, we 
therefore set the burn-in period to be 200 generations.  

The posterior samples for the fitted and derived parameters were made by sampling the chains starting at generation 201 in each run, skipping enough generations to allow the chains
to cross most of the parameter space.
The eight individual posterior samples for each
parameter for each run were 
combined.  

The distributions for the fitted parameters are shown in Fig.~S7, and summarized with the derived parameters in Table~1, which gives the posterior
sample median, the mode (determined by
using 75 bins), and the $1\sigma$ confidence ranges.  For our adopted parameters we
quote the sample medians.  

With the exception of the
inclination $i$ and the Roche lobe
filling factor $f_1$, the 
posterior probability distributions are close to symmetric (Fig.~S7).  
In the case of the filling factor $f_1$, the distribution
is peaked towards the maximum value of 1.0.
The median value of the distribution is about 0.963 and the value
for the best-fitting model is 0.993.  The 95\% confidence lower limit is 0.917.

We find masses of
$40.6^{+7.7}_{-7.1}\,M_{\odot}$ and
$21.2\pm2.2\,M_{\odot}$ (Fig.~S8) for the O-star and
black hole, respectively, as compared to the values of
$19.2\pm 1.9\,M_{\odot}$ and $14.8\pm 1.0\,M_{\odot}$, respectively, from \cite{Orosz2011}.  
The best-fitting radius of the O-star is
$22.3\pm1.8\,R_{\odot}$, which when combined with
its temperature of $T_{\rm eff}=31.1\pm0.7$\,kK,
gives a luminosity of $\log (L/L_{\odot})=
5.63^{+0.07}_{-0.08}$.  Figs.~S9 and S10 show the light curve and radial velocity fits using these median values.  We  find rotation of the donor star consistent with being synchronous with the orbit, unlike the previous solution  \cite{Orosz2011} which had a donor rotating at 1.4 times the orbital frequency.

The strong dependence of the component masses on the distance results from the nearly Roche-lobe filling donor, which is at a known orbital period with well constrained flux and temperature.  The angular size of the donor is set from the combination of temperature and luminosity.  The donor radius is thus proportional to the distance to the system.  Then, since the donor is constrained to be nearly filling its Roche lobe, and the Roche lobe size at a given period is set by the star's mean density, the donor mass scales as its radius cubed, and hence the distance cubed.  This leads to the nearly doubling of the donor mass based on the 20\% increase in distance.

The size of the derived black hole orbit, $a_{\rm BH}=0.160\pm0.013$\,au, corresponds to $73\pm8$\,\micro as at our best-fitting distance of $2.22^{+0.18}_{-0.17}$\,kpc.  This is consistent with the size of the black hole orbit derived independently from our astrometric fit ($58\pm20$\,\micro as), providing confidence in the result. This indicates that the jet cannot undergo a high degree of bending as it propagates outward from the black hole to the radio photosphere, or we would have measured a substantially larger projected orbit when fitting our VLBA astrometric data.

\subsection{Spin Measurement via the Continuum-fitting Method}
\label{sec:spin}

The continuum-fitting method \cite{McClintock2014} is one of the two common techniques used to estimate the dimensionless spin parameter of a black hole, $a_{*}=cJ/GM_{\rm BH}^2$, where $J$ is the black hole angular momentum, $c$ is the speed of light, and $G$ is the gravitational constant. This method relies on determining the inner radius $R_{\rm in}$ of the thin accretion disk, by fitting the continuum X-ray spectrum to the Novikov-Thorne thin disk model~\cite{Novikov1973}. $R_{\rm in}$ is assumed to be located at the innermost stable circular orbit (ISCO) around the black hole, $R_{\rm ISCO}$. $R_{\rm ISCO}$ is directly related to the dimensionless spin parameter $a_{*}$, which for a prograde equatorial orbit decreases from 6 $GM/c^{2}$ to $GM/c^{2}$ as the spin goes from $a_{*}$ = 0 to 1. Fitting the inner radius of the accretion disk then in principle constrains the black hole spin \cite{Bardeen1972}.  However, the spins derived by the continuum-fitting method are sensitive to the black hole mass, the disk inclination, and the source distance, and the uncertainties on these parameters tend to dominate the uncertainties on the derived spins \cite{McClintock2014}.

The continuum fitting method was originally applied only to spectra from the high/soft X-ray spectral state \cite{McClintock2006,Shafee2006}, to ensure that the inner radius of the disk was located at the ISCO, and to avoid confusion introduced by the strongly Comptonized component that is present in the hard and intermediate X-ray spectral states. While Cygnus~\X-1 does not reach the canonical high/soft state, \cite{Steiner2009} have shown that the inner disk radius remains within a few percent of the ISCO for X-ray spectra in which $<25$\% of the thermal seed photons are Compton upscattered.  We applied this criterion to select six archival spectra \cite{Gou2011,Gou2014}, and reanalysed them to constrain the spin parameter.  We used our estimates for the distance, black hole mass and orbital inclination of Cygnus~\X-1. We used Monte Carlo simulations to quantify the uncertainty on the spin introduced by the combined observational uncertainties on $D$, $M$, and $i$ \cite{Gou2014,Miller-Jones2021}.

Assuming that the binary orbital angular momentum vector is aligned with the spin axis of the black hole (as discussed above, and as expected under the evolutionary scenario discussed in the Supplementary text), we find that, consistent with the previous spin results~\cite{Gou2014},  $a_{*} > 0.9985$ at the 3$\sigma$ confidence level.
While the value of $a_{*}$ monotonically decreases with increasing inclination, we find that even given a misalignment angle of $15^\circ$ (equivalent to $i$ = $42.5^\circ$), the fitted spin value would still be as high as $a_{*}=0.9696$. 

While the best-fitting spin is above the maximum value expected from classical modelling \cite{Thorne1974}, this could be attributed to systematic effects in the model arising from factors such as a finite disk thickness, or the inner radius not being exactly at the ISCO.  Regardless, the true spin is likely high, as also found by studies fitting the Fe K$\alpha$ line profile \cite{Duro2016,Tomsick2014,Parker2015,Walton2016,Tomsick2018}.  A detailed analysis of our spin-fitting results and a discussion of the inherent systematics is presented in a companion paper \cite{Zhao2021}.

\subsection{An Updated Mass-Luminosity Relationship}
\label{sec:ML}

We computed a mass-luminosity relation using the Warsaw evolutionary code \cite{Ziolkowski2005,Miller-Jones2021}. We made three changes relative to the model described in \cite{Ziolkowski2014}.  First, we adopted a default effective temperature of 30,500\,K \cite{Shimanskii2012}, which did not lead to any appreciable modification of the mass-luminosity relation.  

Next, we used a different method for choosing the parameter $f_{\rm sw}$ (defined in \cite{Ziolkowski2005} as a simple multiplicative factor applied to the stellar wind mass loss rate prescriptions of \cite{Hurley2000}, to account for uncertainties in our knowledge of the winds in massive stars; see \cite{Higgins2019} for an alternative approach). Two example relations are available \cite{Ziolkowski2014} for fixed values of $f_{\rm sw}$ equal to 2 and 5 (fixed for each mass along the relation). We calibrate the mass-luminosity relation specifically for the mass donor in Cygnus~\X-1. We use the fact that the stellar wind mass outflow rate from the star is known: $\dot{M}_{\rm w} = -(2.57\pm0.05)\times10^{-6}$\,$M_{\odot}$\,yr$^{-1}$ \cite{Gies2003}. While this estimate does not include systematic uncertainties, considerations of wind clumping for comparable late O-type supergiants \cite{Puls2006} suggest that these should be below a factor of two.  For each stellar mass along the mass-luminosity sequence, we adjusted the value of $f_{\rm sw}$ to ensure that when the star evolves to reach the effective temperature $T_{\rm eff} = 30,500$\,K, the calculated mass outflow rate would match the measured value \cite{Gies2003}. The relation was calculated for masses in the range $22$--$47M_{\odot}$. This procedure resulted in values of $f_{\rm sw}$ varying from 4.15 to 0.415 as $\log(M/M_{\odot})$ varied from 1.345 to 1.668. 

Finally, we modified the assumed chemical composition.  For the hydrogen mass fraction and metallicity, we adopted a set of default values \cite{Shimanskii2012} of $X=0.52\pm0.05$ (corresponding to $[{\rm He/H}]=0.42\pm0.05$) and $Z=0.03$ (a super-solar value set by the level of precision available \cite{Shimanskii2012}).  Given the uncertainty on the metallicity, we also computed a relation for $Z=0.02$.  The assumed donor composition was found to have the largest effect on the computed mass-luminosity relation, as shown in Fig.~3.  However, these models do not include the effects of stellar rotation or tidal distortion on temperature and luminosity.

Adopting instead the $1\sigma$ upper bound on the effective temperature found above of $T_{\rm eff} = 31,100$\,K, we found that the predicted luminosity change for a given mass was $<5\%$. To account for systematic uncertainties in the assumed wind mass loss rate, we also tested the effect of changing the wind mass loss rates by 50\%, finding the predicted luminosity change for a given mass to be $<3$\%. Our calculations are therefore insensitive to uncertainties in the wind mass loss rate and effective temperature.

As shown in Fig.~3, our revised parameters of the donor star are consistent with all three of the assumed compositions, within the uncertainties.  Should the ``classical'' composition of $X=0.70$, $Z=0.02$ be correct, then the true hydrogen content would be larger than the measured surface value of $X=0.52$, possibly suggesting enrichment of the donor star during mass transfer (see Supplementary text). 

\section{Supplementary text}

\subsection{Wind mass loss rates}\label{sec:winds}

Wind-driven mass loss rates of massive stars are difficult to accurately measure empirically and to model theoretically (see \cite{Puls:2008,Vink:2017} for reviews).  
We use simplified phenomenological prescriptions with scaling parameters to encode existing uncertainty, following the formalism of \cite{Belczynski2010}.

We focus on two stellar evolution phases that determine the maximum mass of black holes: the luminous blue variable and Wolf-Rayet phases, whose wind mass loss rates are uncertain.  Luminous blue variable winds are conjectured to prevent the star from crossing the Humphreys-Davidson limit \cite{Humphreys1994}.  In our parametrised model, these are independent of metallicity (except indirectly, through conditions determining their onset) and eject mass at a rate 
\begin{equation}
\dot{M}_\mathrm{LBV} = f_\mathrm{LBV} \times 10^{-4} M_\odot\ \mathrm{yr}^{-1}.
\end{equation}
Meanwhile all stars that strip themselves of their envelope through winds beyond the main sequence -- or, in our preferred evolutionary scenario, are stripped by the companion while still on the main sequence -- are subject to strong Wolf-Rayet winds with a metallicity-dependent rate of 
\begin{equation}
\dot{M}_\mathrm{WR} = f_\mathrm{WR} \times 10^{-13} \left(\frac{L}{L_\odot}\right)^{1.5} \left(\frac{Z}{Z_\odot}\right)^{0.86} M_\odot\ \mathrm{yr}^{-1},
\end{equation}
where $L$ is the star's luminosity \cite{Vink2005}.

We use single stellar evolution and wind models as implemented in the binary population synthesis code {\sc compas} \cite{Stevenson:2017,VignaGomez:2018} to explore the impact of Wolf-Rayet and luminous blue variable wind prescriptions on the final black hole mass \cite{Miller-Jones2021}.  We require that black holes with a mass equal to that inferred for Cygnus~\X-1 can be produced at a metallicity of $Z=0.02=1.4 Z_\odot$ (potentially somewhat lower than the metallicity of Cygnus~\X-1, so our wind reduction is conservative \cite{Shimanskii2012}). In Fig.~S11 we show the possible solutions for single stars in the two-dimensional parameter space of $ f_\mathrm{LBV}$ and $f_\mathrm{WR}$.  We find that the mass loss rates of either luminous blue variable winds, or Wolf-Rayet winds, or both, must be reduced relative to the \cite{Belczynski2010} defaults (calibrated to the previous, lower estimates of the Cygnus~\X-1 black hole mass under the assumption that it formed at solar metallicity), which are also used in {\sc compas} \cite{Neijssel2019}.  To form black holes with a mass consistent with that of the black hole in Cygnus~\X-1  from single stars at this metallicity, either the mass loss rates in luminous blue variable winds have to be reduced by a third, or those in Wolf-Rayet winds must be reduced by two thirds.  We also investigated the effect of the standard main sequence winds from O-stars, finding that the effect on the final remnant mass of switching them off entirely was marginal (when compared to the effects of varying either luminous blue variable or Wolf-Rayet winds), and therefore do not consider O-star winds further.

Cygnus~\X-1 is unlikely to contain the highest-mass black hole that can be reached at its metallicity, so this provides only a conservative constraint on the maximum wind mass loss rate.  This simplified analysis assumed single stars; if the progenitor of the Cygnus~\X-1 black hole was stripped of its hydrogen envelope during the main sequence as conjectured below, Wolf-Rayet winds will kick in earlier, again making these constraints conservative.  Regardless, even the moderate wind strength reduction assumed here has consequences for the mass distribution of compact objects \cite{Mennekens2014, Barrett:2017FIM} and supernova models.  For example, \cite{Renzo:2017} find that varying wind assumptions, in addition to affecting the core mass, substantially impact the compactness parameter of stellar cores, which is connected with the `explodability' in a supernova.  Reduced winds would also impact the enrichment of the interstellar medium by massive stars \cite{Maeder:1983}, and their role in the re-ionization of the universe; while ionising radiation from massive stars made the universe transparent \cite{HaimanLoeb:1997}, it strongly depends on stripping by winds or binary interactions \cite{Gotberg:2017}.

\subsection{Comparison to other systems}

Dynamical black hole mass measurements of the high-mass X-ray binaries IC10~\X-1 \cite{Prestwich2007,Silverman2008} and NGC~300~\X-1 \cite{Carpano2007,Crowther2010,Binder2011}
favour higher masses than our revised measurement of the Cygnus~\X-1 black hole.  However, the source of the emission lines used for those measurements is unknown.  For example \cite{Laycock2015,Binder2015} point out that in the case of Wolf-Rayet companions, the emission lines likely originate in the wind and thus indicate the wind velocity rather than the binary orbital velocity, which would invalidate the black hole mass measurements.  IC10~\X-1 and NGC~300~\X-1 reside in galaxies with sub-solar metallicities, so the evolution of their progenitors is likely to differ from that of the black hole progenitor in Cygnus~\X-1.  

LB-1 (LS V +22 25) in the Milky Way has been suggested to contain a $70M_{\odot}$ black hole \cite{Liu2019}.  However, other work has challenged the interpretation of the broad H$\alpha$ line used to determine the black hole mass, and suggested that the mass of the unobserved companion (which may not even be a black hole; \cite{Shenar:2020}) is closer to $5M_{\odot}$ \cite{AbdulMasih2019,Eldridge2019,Elbadry2019,Shenar:2020}.

\subsection{Formation history}\label{sec:history}

We investigate a possible formation channel for Cygnus~\X-1.
The close separation indicates that the binary must have interacted, with mass transfer from the progenitor of the black hole, the primary, onto its companion, the secondary.  Such mass transfer would typically remove the envelope of the progenitor, which would dominate its moment of inertia and, assuming reasonably efficient angular momentum transport within the star, would contain the bulk of the angular momentum.  Therefore, removing this envelope would leave behind a slowly spinning black hole \cite{Kushnir:2016,Zaldarriaga:2017,HotokezakaPiran:2017,Bavera:2019,FullerMa:2019,Belczynski:2020}. 

However, observations point to a rapidly rotating black hole, aligned with the orbital angular momentum \cite{SuppMaterial}.  The black hole must therefore retain its birth spin, because the mass of a black hole would have to be doubled in order to appreciably change spin after birth \cite{Thorne1974}, which is inconsistent with the short lifetime of the system \cite{Axelsson:2011}; even Eddington-limited accretion over the entire 4\,Myr lifetime of the donor (both assumptions that are likely to be substantially overestimated) would only spin up the black hole to a maximum of $a_{\ast}=0.3$.  The observed separation is too large for tides to spin up the stellar core of the primary after its envelope was removed.   Chemically homogeneous evolution \cite{MandeldeMink:2016,Marchant:2016} could be responsible for the rapid spin of the black hole; however, given the observed mass ratio, it appears unlikely that the primary evolved chemically homogeneously but the secondary did not, as evidenced by its factor of $\sim 2$ expansion from its zero-age main sequence radius.
Spin-up is possible during the supernova itself through tidal torquing by the companion of ejecta which eventually fall back \cite{Schroeder:2018}, but this generally yields lower spins than complete fallback.

A channel for forming a rapidly rotating black hole in a close binary was proposed by \cite{Qin2019}: gradual case A mass transfer from the late main-sequence primary allows it to be spun up through tides while removing enough hydrogen-rich material to prevent post-main sequence expansion and loss of angular momentum through envelope stripping.  While the details are uncertain, this channel appears to explain the similar properties of the high-mass X-ray binaries Cygnus~\X-1, LMC~\X-1, and M33~\X-7, including orbital separations of $\sim 50 R_\odot$, and high spins \cite{MillerMiller:2015} (see \cite{Podsiadlowski:2003,Valsecchi:2010,Wong:2012} for previous evolutionary studies of these systems).

Assuming evolution through this channel, the primary would have lost some mass through winds during the Wolf-Rayet phase, widening the binary.  However, this mass loss cannot have been too great, due to the mass of the black-hole remnant, and to avoid excessive spin-down. Eventually, the primary collapsed into an aligned, rapidly spinning black hole, likely through nearly complete fallback with little to no natal kick \cite{Fryer:2012}, consistent with the few km\,s$^{-1}$ drift velocity of the binary relative to its Cygnus OB3 birth association \cite{Mirabel2003}.  The lack of substantial mass loss and natal kick is supported by the very low binary eccentricity \cite{SuppMaterial}, although the fact that the eccentricity is inconsistent with zero may argue for a minimal mass loss of $\sim 1 M_\odot$ (perhaps through neutrino emission) and imperfect tidal circularisation through tides operating on the expanding secondary.

The combination of case A mass transfer and Wolf-Rayet winds is presumably responsible for the unusual surface abundances of the secondary, including surface helium abundance enhanced by more than a factor of two \cite{Shimanskii2012}.  
This is different from the model of \cite{Podsiadlowski:2003}, who assumed that the surface abundances are due to partial stripping of the secondary during a previous Roche lobe overflow episode onto the black hole; however, in their model the secondary is now less massive than the black hole, which is inconsistent with the present measurements. Enrichment during the supernova itself may have occurred in the black hole X-ray binary GRO J1655-40, rather than during previous mass transfer \cite{Israelian:1999,Podsiadlowski:2002SN}.  But as discussed above we favour nearly complete fallback in Cygnus~\X-1.  Moreover, if the current secondary were, in fact, a partially stripped evolved star, we would expect it to have a higher core mass and therefore a higher luminosity than a normal main sequence star of its current mass, which is not supported by observations (Fig.~3). 

Our scenario relies on a small quantity of enriched material being responsible for the observed secondary surface abundance.  This demands that the material accreted from the primary is retained in a thin surface layer on the secondary, rather than being mixed throughout the star.  Late main-sequence stars with the mass and metallicity of the secondary should only have a very thin convective surface layer, preventing efficient mixing (although the impact of tides and rotation could enhance convection).  While temperature inversion in the accreted material would suppress the Rayleigh-Taylor instability, the more massive helium-rich material would be susceptible to thermohaline mixing \cite{Kippenhahn:1980,BraunLanger:1995}, which would eventually distribute the accreted material through the envelope, reducing the observed surface abundance.  Therefore, the observed enhancement points to fairly recent enrichment and collapse of the black hole, likely $\lesssim 10^5$ years ago.  This requirement is also necessary to avoid the enriched material being blown off by stellar winds from the secondary.  This timescale is consistent with the few $\times 10^4$ yr estimate for the age of the jet \cite{Russell:2007}, and hence the age of the black hole if the jet switched on promptly after the black hole was formed, as expected for wind mass transfer  given the proximity of the two components.  

The formation scenario and its inherent uncertainties are discussed in detail elsewhere \cite{Neijssel:2020CygX1}.

\clearpage

\renewcommand{\thefigure}{{\bf S1}}
\begin{figure}
\centering
\includegraphics[width=\textwidth,angle=0]{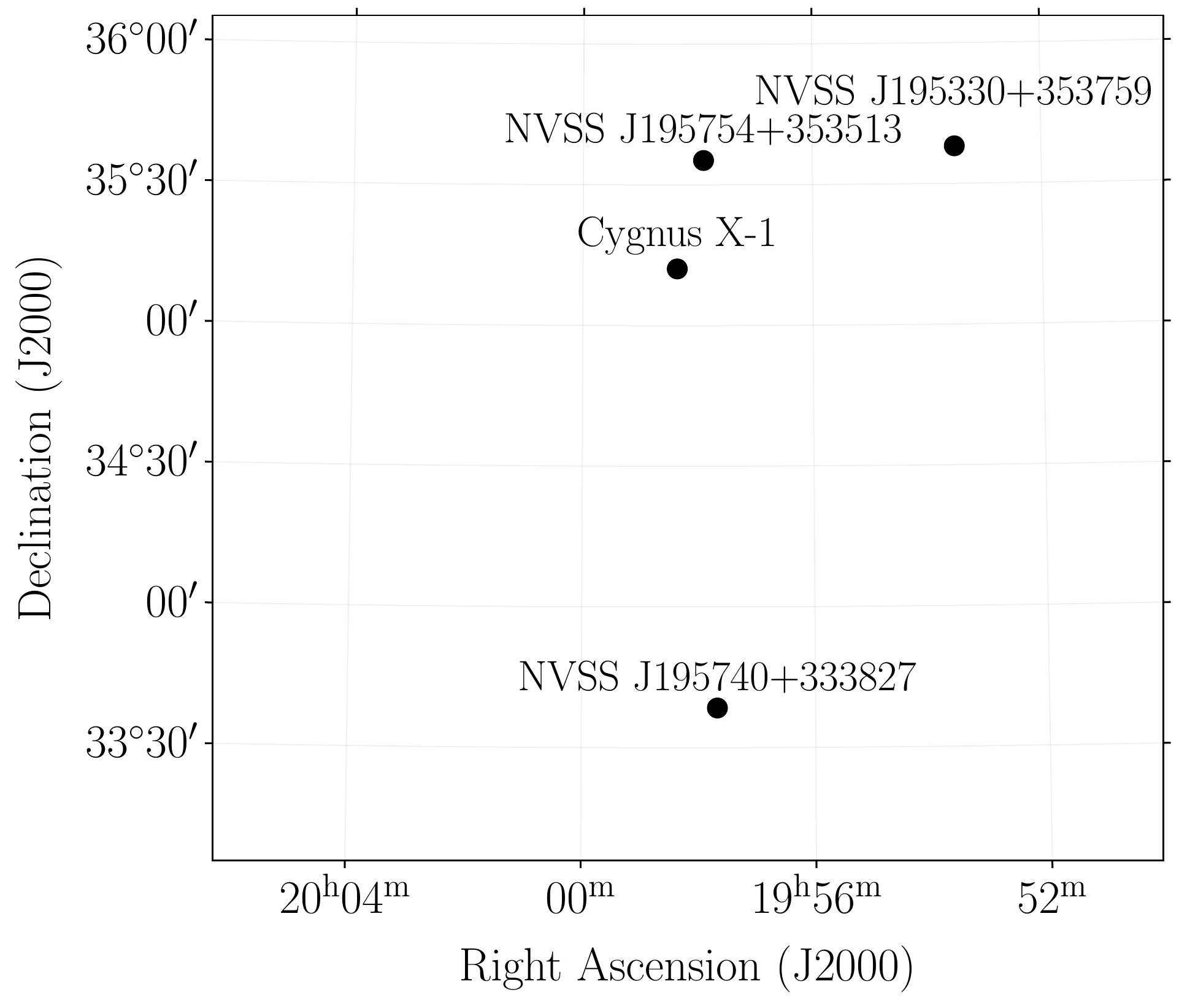}
\caption{{\bf Locations of our observed sources on the plane of the sky.} We show the positions of our calibrator sources, NVSS~J195330+353759 (J1953+3537), NVSS~J195754+353513 (J1957+3535), and NVSS~J195740+333827 (J1957+3338), as well as that of Cygnus~\X-1.  Horizontal and vertical axes are Right Ascension and Declination, respectively.  North is upwards, and east is to the left.  The calibrator throw from J1957+3338 is similar to both Cygnus~\X-1 and the check source, J1957+3535, constraining the systematic uncertainties affecting our measurements of Cygnus~\X-1. All coordinates are in the International Celestial Reference System (ICRS).\label{fig:cx1_field}}
\end{figure}

\renewcommand{\thefigure}{{\bf S2}}
\begin{figure}
\centering
\includegraphics[width=0.95\textwidth]{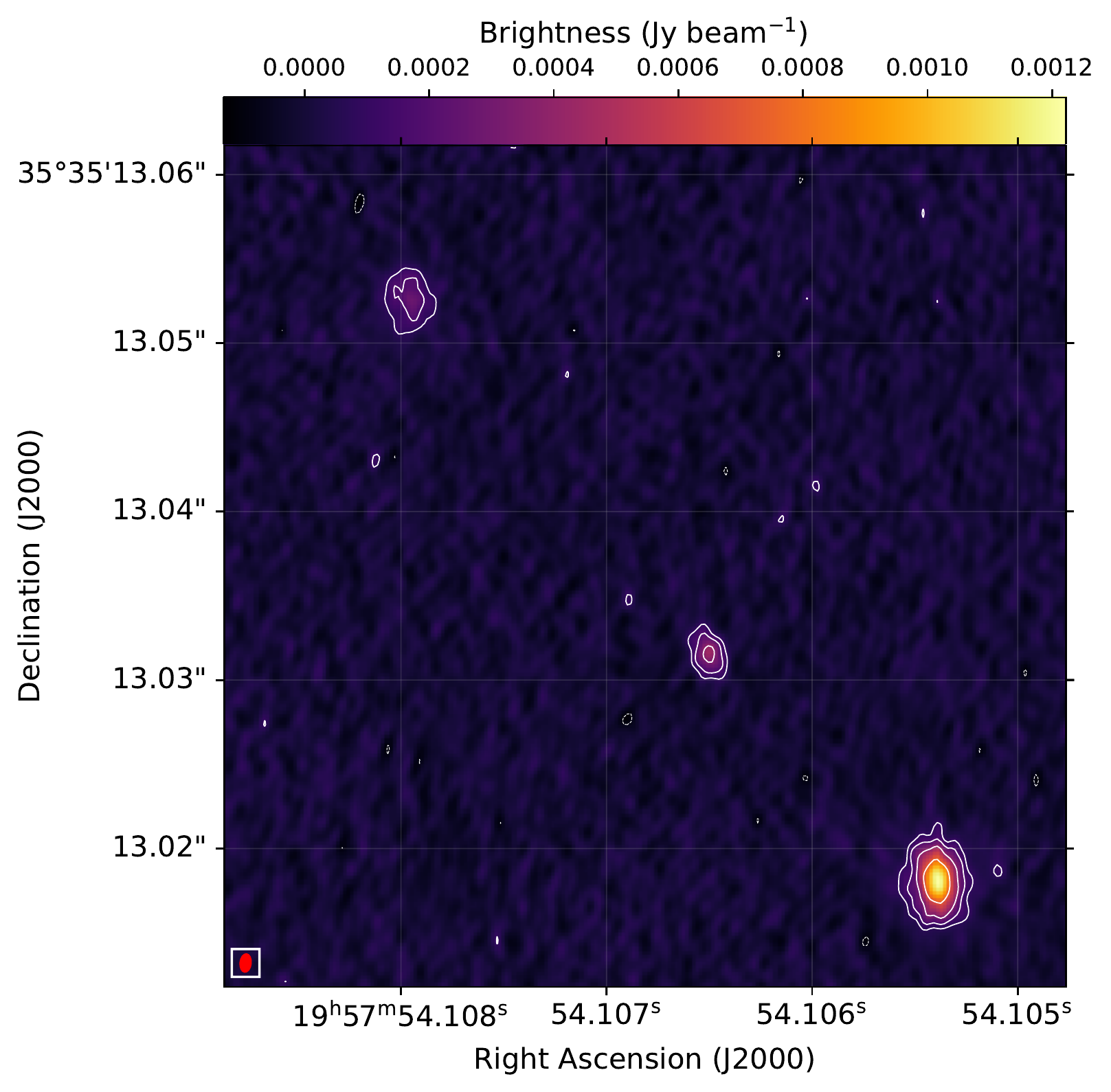}
\caption{{\bf 8.4-GHz VLBA image of J1957+3535.} The image was made from a stack of all six epochs of data from 2016 May and June.  The source is comprised of three components, all slightly extended relative to the VLBA beam.  Contours are at levels of $\pm 2^n$ times 90\,\micro Jy\,beam$^{-1}$, where $n=0,1,2,...$  The rms noise level in the image is 26\,\micro Jy\,beam$^{-1}$.  The fitted locations of the central component have a standard deviation of 14 and 30\,microarcseconds (\micro as) in R.A.\ and Dec., respectively, over our six epochs of observation. The red ellipse in the bottom left corner represents the synthesised beam. \label{fig:j1957+3535_stack}}
\end{figure}

\renewcommand{\thefigure}{{\bf S3}}
\begin{figure}
        \centering
        \includegraphics[width=\textwidth]{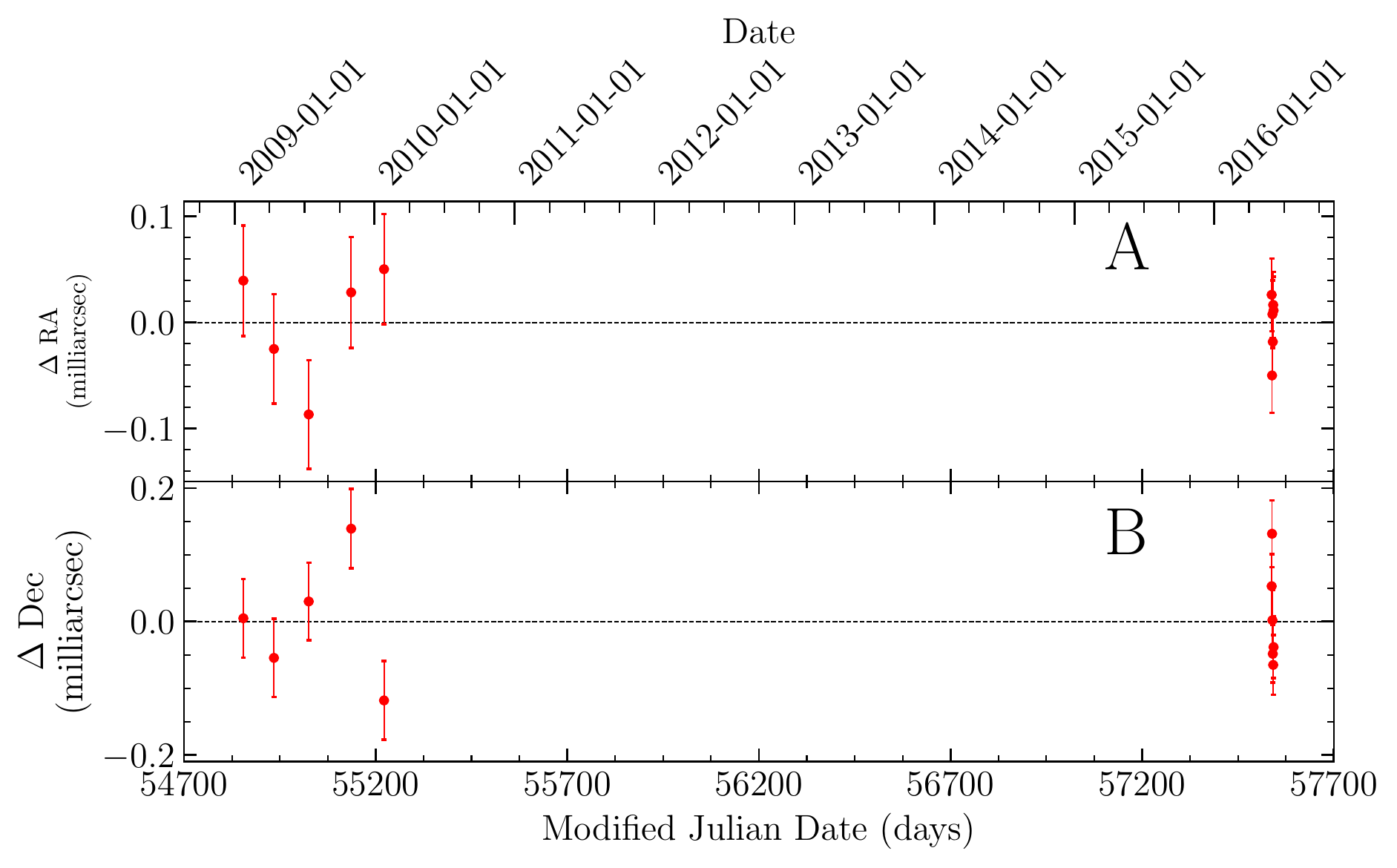} 
        \caption{{\bf Residuals from the two-dimensional astrometric fit.} Our model incorporated proper motion, parallax, and orbital motion, fitted in both Right Ascension (A) and Declination (B).  Residuals exceed 0.1\,mas, substantially larger than we expected from our estimated systematic and statistical uncertainties, which are reflected in the sizes of our error bars, shown at the 68\% confidence level.} \label{fig:model_residuals}
\end{figure}

\renewcommand{\thefigure}{{\bf S4}}
\begin{figure}
\centering
\includegraphics[width=\textwidth]{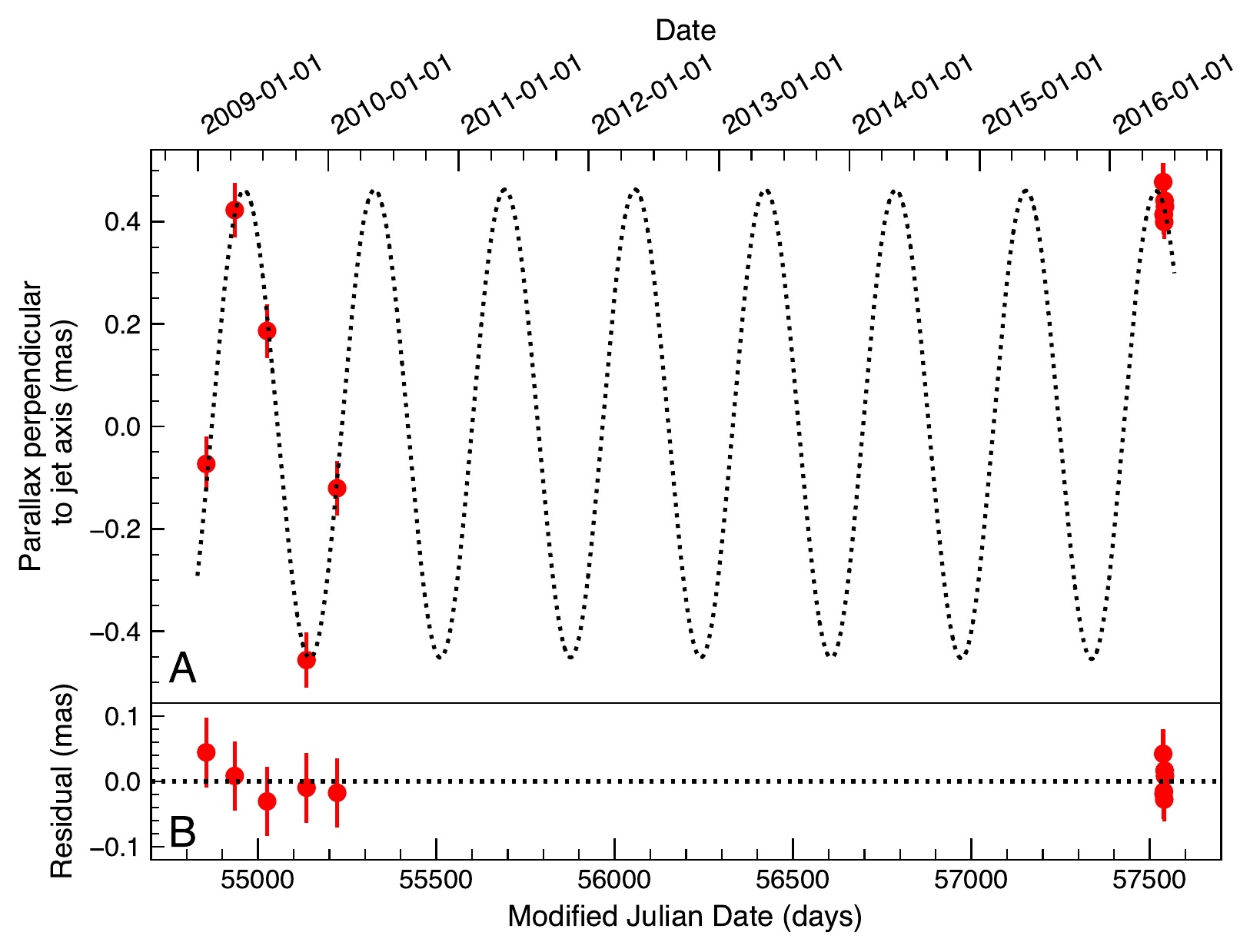}
\caption{{\bf The best-fitting parallax signal from our one-dimensional model fitting.} We fitted the data perpendicular to the jet axis on the plane of the sky. (A) The parallax signal after removing the reference position, proper motion and orbital signatures. (B) The residuals after subtracting the best-fitting parallax of $\pi=458\pm35$\,\micro as.  Error bars are shown at the 68\% confidence level.\label{fig:1dparfit}}
\end{figure}

\renewcommand{\thefigure}{{\bf S5}}
\begin{figure}
\vspace{-1cm}
\centering
\includegraphics[width=\textwidth]{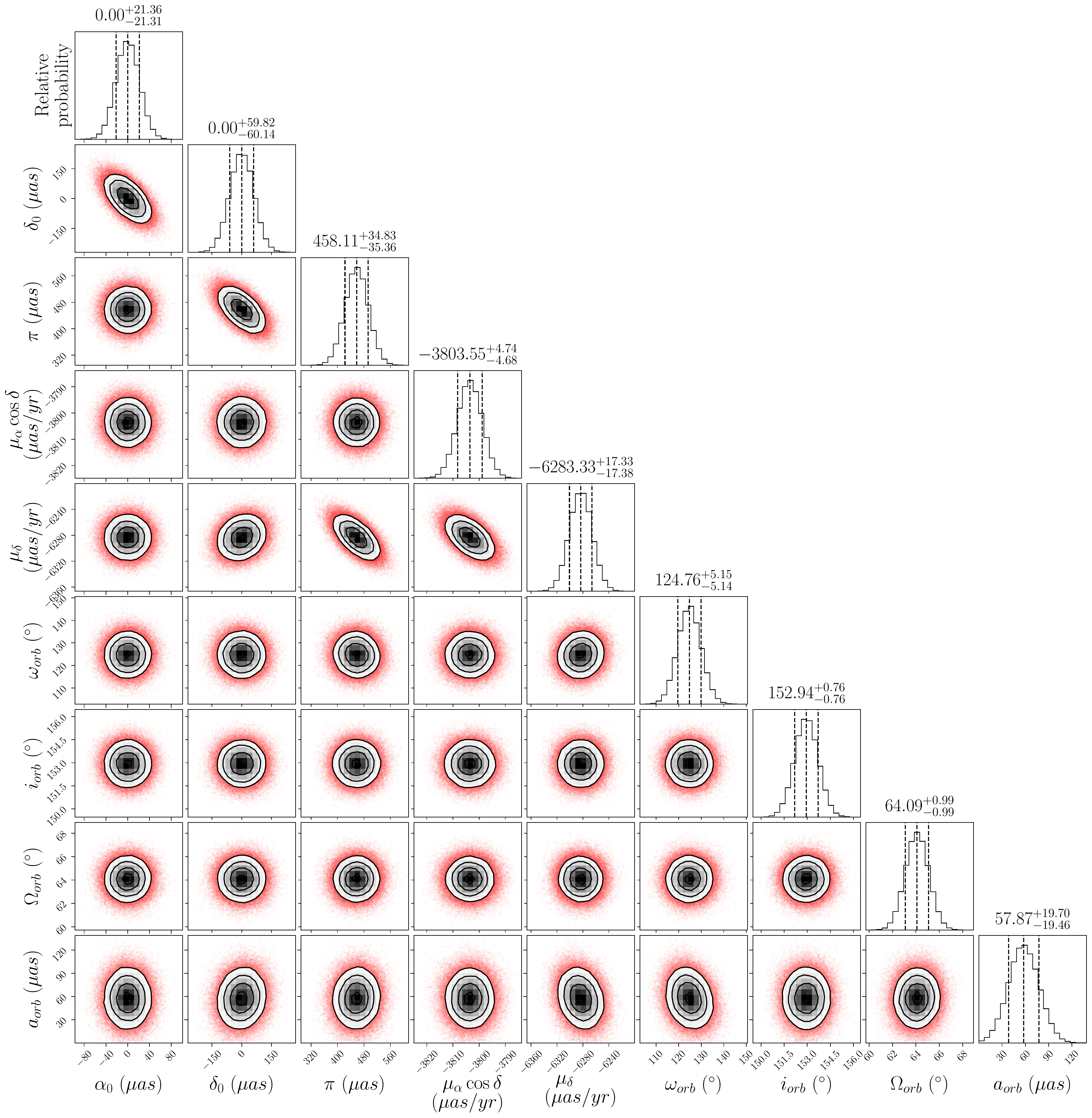}
\caption{{\bf Results of our PyMC astrometric model fitting for Cygnus~\X-1.}  We fitted the position ($\alpha_0$, $\delta_0$; relative to the best-fitting position, in \micro as), proper motions ($\mu_{\alpha}\cos\delta$, $\mu_{\delta}$; in \micro as\,yr$^{-1}$), parallax ($\pi$, in \micro as), inclination angle of the orbit ($i$, in degrees), argument of periastron ($\omega$, in degrees), longitude of the ascending node ($\Omega$, in degrees), and semi-major axis of the black hole orbit ($a_{\rm BH}$, in \micro as).  Histograms show the one-dimensional posterior probability distributions, and contour plots show the two-dimensional posterior probability distributions from our nine-dimensional parameter space.  This shows both the spread of results and their covariances. The dashed vertical lines in the histograms represent the 1$\sigma$ credible intervals, and the contour lines in the contour plots show the 1, 2 and 3$\sigma$ regions (from innermost to outermost, respectively). The red dots in the contour plots represent posterior realizations outside the 3$\sigma$ contour.\label{fig:corner}}
\end{figure}

\renewcommand{\thefigure}{{\bf S6}}
\begin{figure}
\centering
\includegraphics[width=0.8\textwidth]{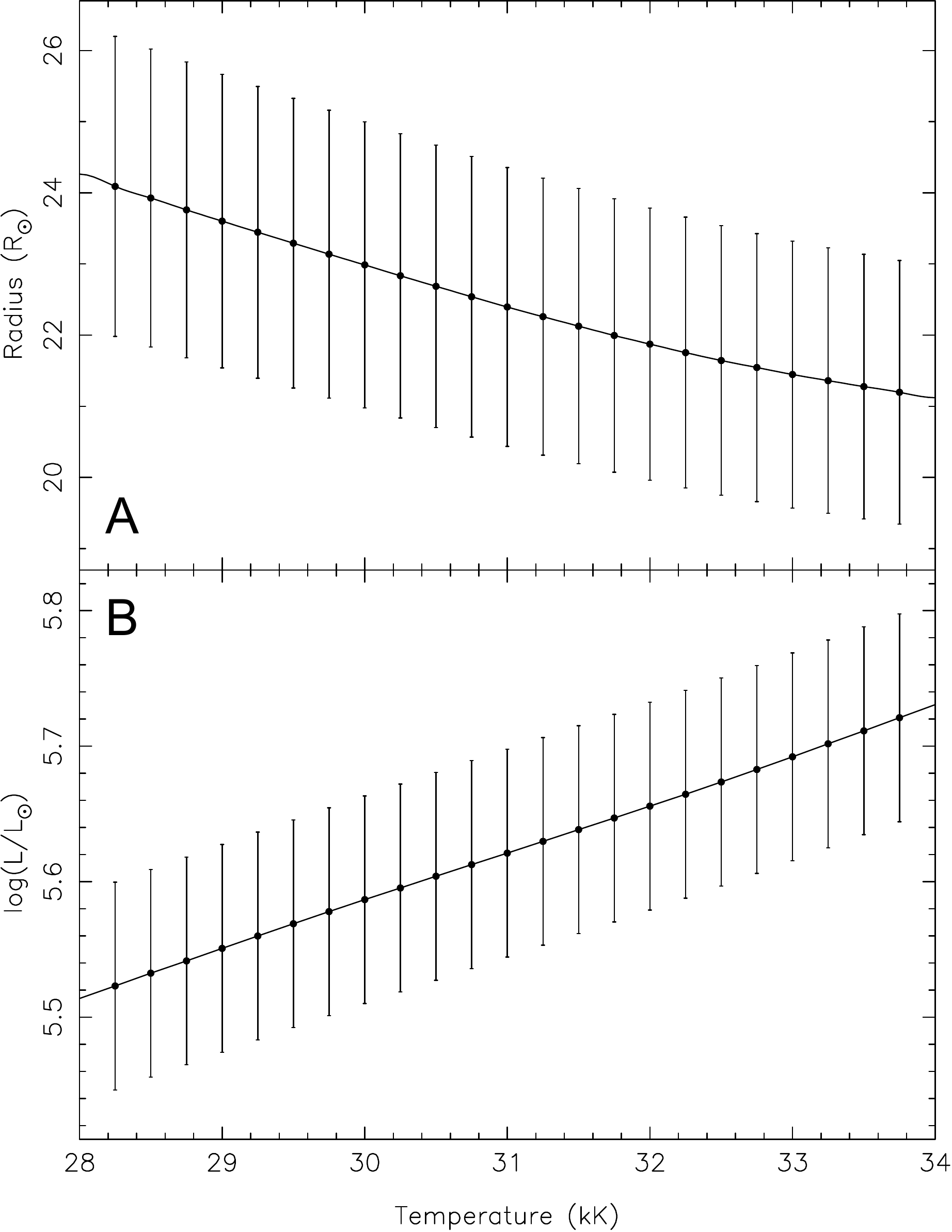}
\caption[test]{{\bf Stellar parameters of Cygnus~\X-1 as a function of its effective temperature.}  We show both the radius (A) and luminosity (B) of the O-star.  The lines joining the points are not fits, and are shown only to highlight the broad trends in the data.  The dependence of the bolometric corrections (see \cite{Orosz2011}) on temperature leads to the luminosity variation.\label{fig:plotrad}}
\end{figure}

\renewcommand{\thefigure}{{\bf S7}}
\begin{figure}
\centering
\includegraphics[height=0.83\textwidth]{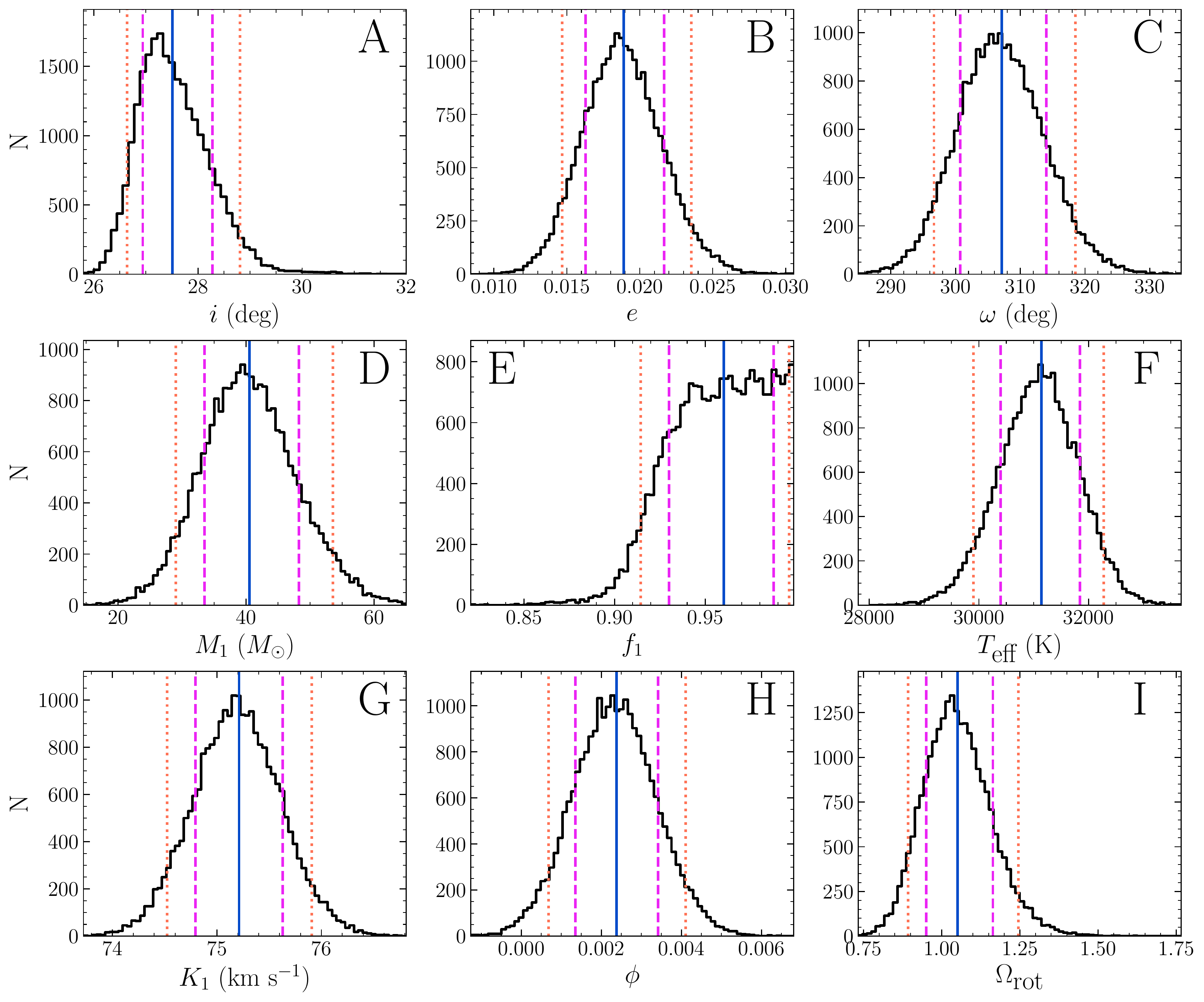}
\caption{{\bf Posterior probability distributions for all nine parameters from our optical model fitting.} $N$ is the number of posterior samples in a given bin, out of a total of 24,360.  The only two non-symmetric distributions are for the orbital inclination angle $i$ (A), and the Roche Lobe filling factor for the secondary, $f_{1}$ (E). The inclination angle is well-constrained, and the secondary appears very close to Roche lobe filling.  We are only able to place a 95\% confidence lower limit on the filling factor, of $>0.914$. Blue solid line shows the sample medians, and pink dashed lines show the lower and upper 15.87\% points, which between them encompass the $1\sigma$ parameter ranges. These values are listed in Table~1.  Orange dotted lines show the 90\% Bayesian credible intervals.\label{fig:histfig01}}
\end{figure}

\renewcommand{\thefigure}{{\bf S8}}
\begin{figure}
    \centering
    \includegraphics[height=0.7\textwidth]{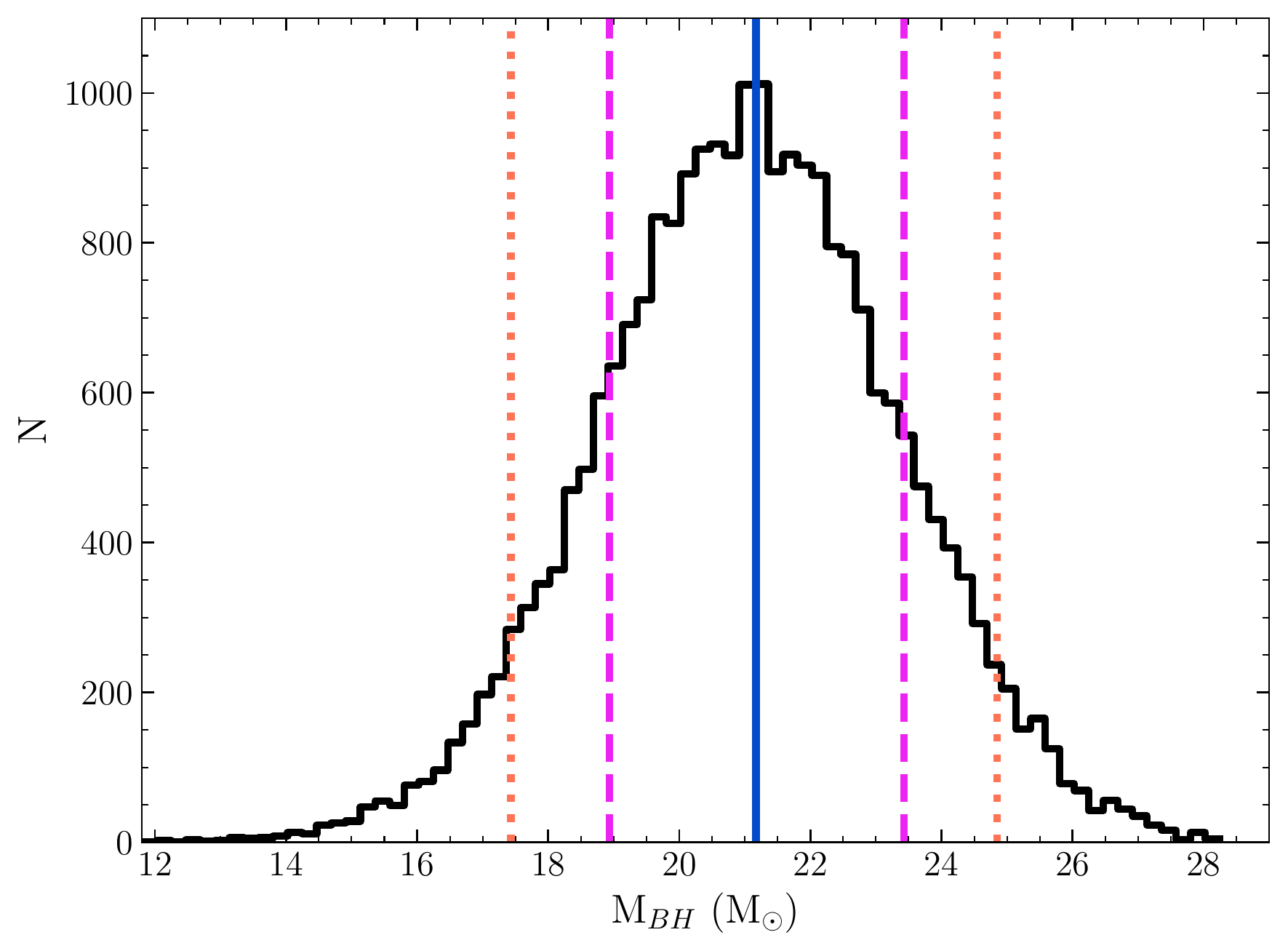}
    \caption{{\bf Posterior probability distribution of the black hole mass determined from our optical light curve and radial velocity modelling.} $N$ is the number of posterior samples in a given bin, out of a total of 24,360.  Blue solid line shows the sample median, and pink dashed lines show the lower and upper 15.87\% points, which between them encompass the $1\sigma$ range for the black hole mass, of 18.9--23.4$M_{\odot}$.  Orange dotted lines show the 90\% Bayesian credible interval of 17.4--24.8$M_{\odot}$. \label{fig:Mbh}}
\end{figure}

\renewcommand{\thefigure}{{\bf S9}}
\begin{figure}
\centering
\includegraphics[width=\textwidth]{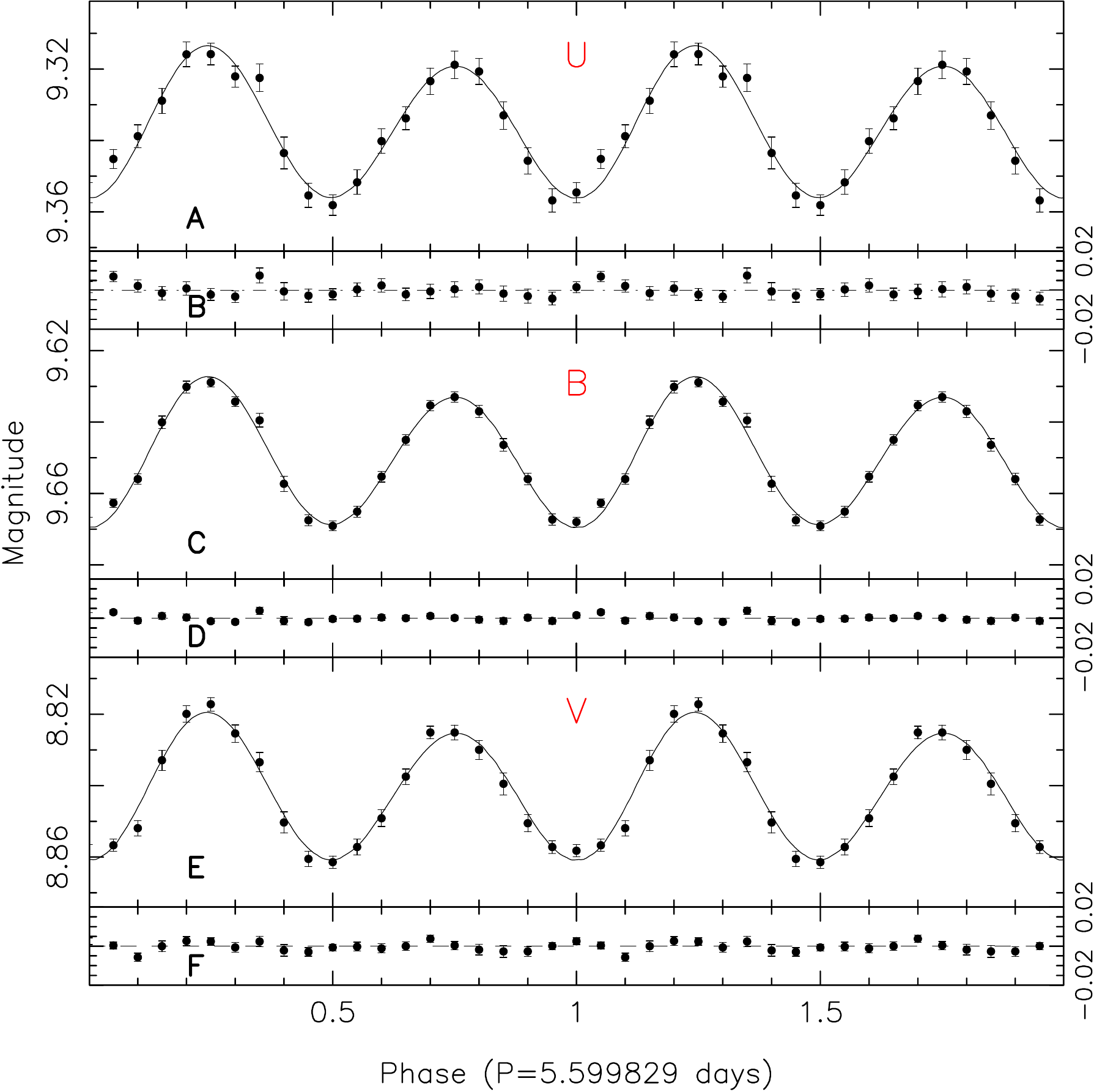}
\caption{{\bf Light curve results from our \textbf{\textsc{elc}} model fitting}.  (A) The $U$-band light curve and the best-fitting model. (B) The $U$-band residuals. (C) The $B$-band light curve and the best-fitting model. (D) The $B$-band residuals. (E) The $V$-band light curve and the best-fitting model. (F) The $V$-band residuals.  Error bars are shown at the 68\% confidence level.  The data are duplicated over two orbital cycles for clarity. \label{fig:ubvfitfig}}
\end{figure}

\renewcommand{\thefigure}{{\bf S10}}
\begin{figure}
\centering
\includegraphics[width=\textwidth]{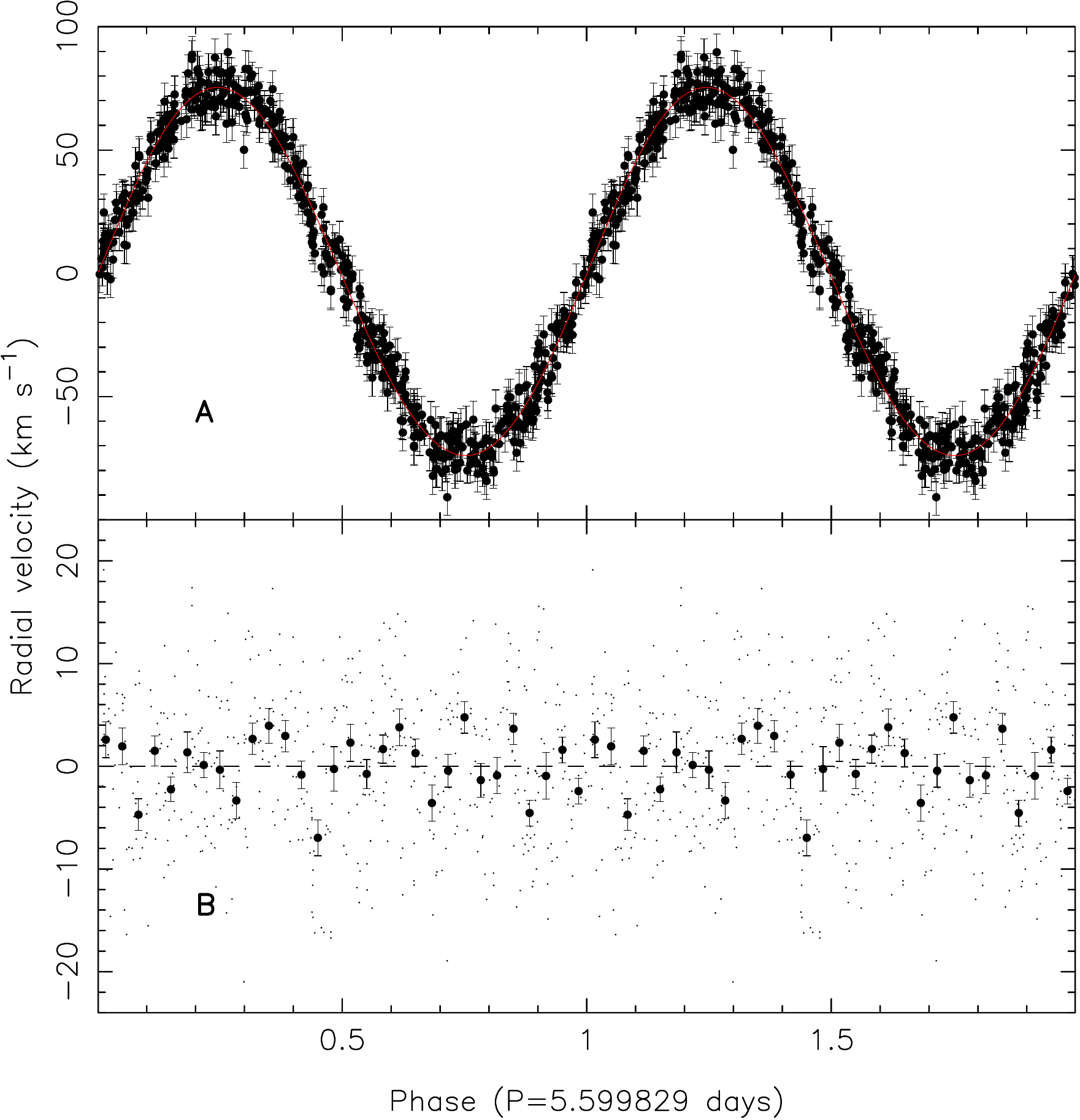}
\caption{{\bf Radial velocity results from our \textbf{\textsc{elc}} model fitting.}  (A) The phase-folded radial velocity measurements (data points) and the best-fitting model (red line). (B) The radial velocity residuals (dots).  The filled circles are the averaged residuals in 30 bins, with uncertainties representing the standard deviation within each bin.  For context, the fitted semi-amplitude of the radial velocity curve is $75.2\pm0.4$\,km\,s$^{-1}$. Data are duplicated over two orbital cycles for clarity. \label{fig:rvfitfig}}
\end{figure}

\renewcommand{\thefigure}{{\bf S11}}
\begin{figure}
\centering
\includegraphics[width=0.75\textwidth]{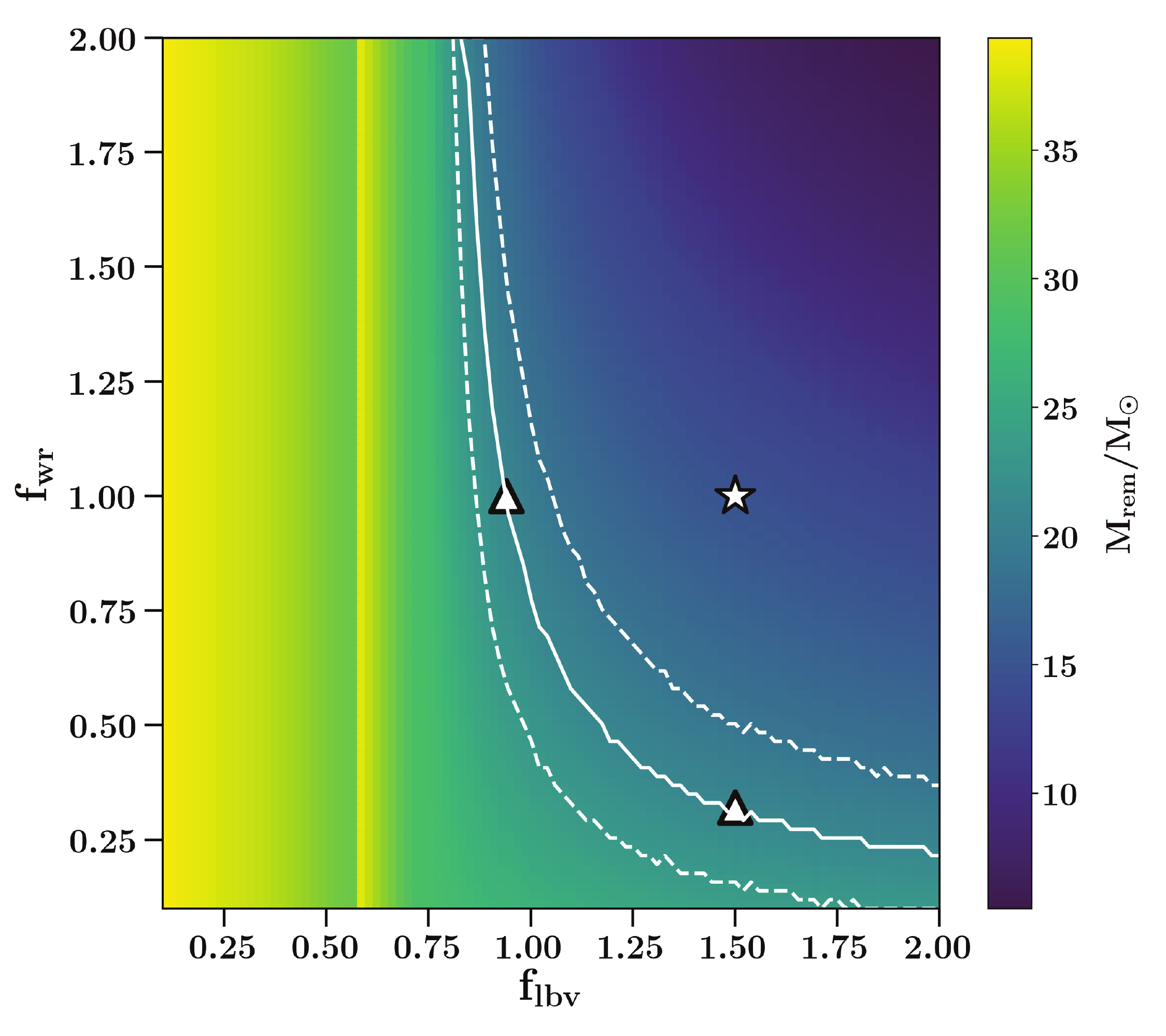}
\caption[test]{{\bf The inferred wind strength in Cygnus~\X-1, relative to standard prescriptions.} We show the maximum remnant mass from the evolution of single stars at $Z=0.02$ as calculated using {\sc compas}, as a function of two parameters describing the strength of luminous blue variable winds, $f_\mathrm{lbv}$, and Wolf-Rayet winds, $f_\mathrm{wr}$.  The solid and dashed white lines indicate the median values and 68\% confidence uncertainty on our Cygnus~\X-1 mass measurements.  
The star denotes default choices calibrated to the previous Cygnus~\X-1 mass measurement 
\protect{\cite{Belczynski2010}}. Parameter choices all fall below and to the left of it. Triangles show the isolated effects of reducing the mass loss rates in Wolf-Rayet winds by a factor of three, and reducing those in luminous blue variable winds by a third.
\label{fig:BHmass}}
\end{figure}

\clearpage

\renewcommand{\thetable}{{\bf S1}}
\begin{table}
 \centering
 \caption[test]{{\bf Measured positions of the radio core of Cygnus~\X-1.} Our images were fitted with an elliptical Gaussian function in the image plane (appropriately corrected for the measured offsets of the secondary phase reference calibrator for epoochs F--K).  Tabulated uncertainties include both statistical and systematic uncertainties, as described in \cite{SuppMaterial}.  Epochs A--E were taken in 2009--2010 under proposal code BR141 \protect{\cite{Reid2011}}, and epochs F--K are our VLBA measurements from 2016, taken under proposal code BM429.}
\label{table:obs_positions}
 \begin{tabular}{l c c c c c} 
 \hline
 Epoch & Modified & Right Ascension & Uncertainty & Declination & Uncertainty \\
 & Julian Date & & & & \\
 & (days) & & (\micro s) & & (\micro as)\\
 \hline
A & 54854.793 & 19$^{\rm h}$58$^{\rm m}$21\fs6729257 & 4.3 & 35$^{\circ}$12$^{\prime}$5\farcs728295 & 59\\
B & 54934.574 & 19$^{\rm h}$58$^{\rm m}$21\fs6728871 & 4.2 & 35$^{\circ}$12$^{\prime}$5\farcs727202 & 59\\
C & 55025.326 & 19$^{\rm h}$58$^{\rm m}$21\fs6727746 & 4.2 & 35$^{\circ}$12$^{\prime}$5\farcs726103 & 58\\
D & 55136.023 & 19$^{\rm h}$58$^{\rm m}$21\fs6726354 & 4.3 & 35$^{\circ}$12$^{\prime}$5\farcs723756 & 60\\
E & 55221.788 & 19$^{\rm h}$58$^{\rm m}$21\fs6726232 & 4.2 & 35$^{\circ}$12$^{\prime}$5\farcs721701 & 59\\
F & 57537.437 & 19$^{\rm h}$58$^{\rm m}$21\fs6706713 & 2.8 & 35$^{\circ}$12$^{\prime}$5\farcs682730 & 48\\
G & 57538.434 & 19$^{\rm h}$58$^{\rm m}$21\fs6706620 & 2.9 & 35$^{\circ}$12$^{\prime}$5\farcs682710 & 50\\
H & 57539.432 & 19$^{\rm h}$58$^{\rm m}$21\fs6706713 & 2.6 & 35$^{\circ}$12$^{\prime}$5\farcs682521 & 46\\
I & 57540.429 & 19$^{\rm h}$58$^{\rm m}$21\fs6706746 & 2.5 & 35$^{\circ}$12$^{\prime}$5\farcs682501 & 43\\
J & 57541.426 & 19$^{\rm h}$58$^{\rm m}$21\fs6706764 & 2.6 & 35$^{\circ}$12$^{\prime}$5\farcs682557 & 45\\
K & 57542.424 & 19$^{\rm h}$58$^{\rm m}$21\fs6706679 & 2.6 & 35$^{\circ}$12$^{\prime}$5\farcs682604 & 47\\
\hline
\end{tabular}
\end{table}

\renewcommand{\thetable}{{\bf S2}}
\begin{table*}
\centering
\caption{{\bf MCMC priors adopted for the two-dimensional astrometric model fitting.} $(\alpha_0,\delta_0)$ are the reference positions in RA and Dec.\ on the reference date, MJD 56198.0.  $\mathcal{U}$ denotes a uniform distribution with the given range, and $\mathcal{N}$ denotes a normal (Gaussian) distribution with mean $\mu$ and standard deviation $\sigma$.
\label{tab:priors}}
\begin{threeparttable}
    \centering
 \begin{tabular}{c c c c}
 \hline
 Parameter & Description & Prior distribution & Units \\
 \hline
 $\alpha_0$ & R.A.\ reference position & $\mathcal{U}$($-0.2, 0.2$)\tnote{$^{\ast}$} & arcseconds\\
 $\delta_0$ & Dec.\ reference position & $\mathcal{U}$($-0.2, 0.2$)\tnote{$^{\ast}$} & arcseconds\\
 $\mu_{\alpha}\cos\delta$ & R.A.\ proper motion & $\mathcal{U}$($-20, 20$) & mas\,yr$^{-1}$\\
 $\mu_{\delta}$ & Dec.\ proper motion & $\mathcal{U}$($-20, 20$) & mas\,yr$^{-1}$\\
 $\pi$ & Parallax & $\mathcal{U}$($0.1, 0.9$) & mas\\
 $i$ & Orbital inclination & $\mathcal{N}$($\mu=152.94$, $\sigma=0.76$)\tnote{$^{\dagger}$} & degrees\\
 $\omega$ & Argument of periastron & $\mathcal{N}$($\mu=127.6$, $\sigma=5.3$)\tnote{$^{\dagger}$} & degrees\\
 $\Omega$ & Longitude of ascending node & $\mathcal{N}$($\mu=64.0$, $\sigma=1.0$)\tnote{$^{\ddagger}$} & degrees\\
 $a_{\rm BH}$ & BH orbital semimajor axis & $\mathcal{U}$($0.0, 0.5$) & mas\\
\hline
\end{tabular}
\begin{tablenotes}
\item[$^{\ast}$] Relative to mean of measured positions in R.A.\ and Dec..
\item[$^{\dagger}$] Taken from optical fitting results of \cite{Orosz2011}.
\item[$^{\ddagger}$] Assuming jet axis and orbital angular momentum are aligned.
\end{tablenotes}
\end{threeparttable}
\end{table*}

\renewcommand{\thetable}{{\bf S3}}
\begin{table*}
\centering
\caption{{\bf Astrometric and orbital parameters of Cygnus~\X-1 from our one-dimensional MCMC model fitting.} We used the priors described in Table~S2.  $(\alpha_0,\delta_0)$ are the reference positions in RA and Dec.\ on the reference date, MJD 56198.0. The stated value is the median of the posterior distribution, with the $1\sigma$ uncertainty (in brackets, applied to the final digits) defined as the spread between the median and percentiles 15.9 and 84.1 of the posterior probability distribution.  The final two columns show the 90\% Bayesian credible interval (the 5th and 95th percentiles of the posterior distribution) for the 1-D fit perpendicular to the jet axis. 
\label{tab:fit_results}}
\begin{threeparttable}
    \centering
 \begin{tabular}{l c c c c}
 \hline
 Parameter & 2-D fit\tnote{$^{\ast}$} & 1-D fit\tnote{$^{\dagger}$} & 1-D fit\tnote{$^{\dagger}$} & 1-D fit\tnote{$^{\dagger}$} \\
 & & & 5th percentile & 95th percentile\\
 \hline
$\alpha_0$ (19$^{\rm h}$58$^{\rm m}$)\tnote{$^{\ddagger}$}     &  21\fs6717793(17) & 21\fs6717793(17) & 21\fs6717770 & 21\fs6717817\\
$\delta_0$ (35\degr\ 12$^{\prime}$)\tnote{$^{\ddagger}$}    &   5\farcs705435(18) & 5\farcs705497(60) & 5\farcs705399 & 5\farcs705597\\
$\mu_{\alpha}\cos\delta$ \,(mas\,yr$^{-1}$)   &  $-3.804(5)$ & $-3.804(5)$ & $-3.801$ & $-3.796$ \\
$\mu_{\delta}$\,(mas\,yr$^{-1}$)    & $-6.312(6)$ &   $-6.283(17)$ & $-6.312$ & $-6.254$ \\
$\pi$\,(mas)   & $0.535(28)$ & $0.458(35)$ & $0.399$ & $0.516$\\
$i$ \,($^{\circ}$)\tnote{$^{\mathsection}$}   &  $153.0(8)$ &  $152.9(7)$ & $151.7$ & $154.2$\\
$\omega$ \,($^{\circ}$) &  $118(5)$ &  $125(5)$ & $116$ & $133$\\
$\Omega$  \,($^{\circ}$) &  $64.4(1.0)$ &  $64.1(1.0)$ & $62.4$ & $65.7$ \\
$a_{\rm BH}$ \,(\micro as) & $89(15)$ & $58(20)$ & $25$ & $90$ \\
\hline
\end{tabular}
\begin{tablenotes}
\item[$^{\ast}$] Two-dimensional fit to the measured Right Ascension and Declination values, as described in \cite{SuppMaterial}.
\item[$^{\dagger}$] One-dimensional fit perpendicular to the jet axis, adopting priors for $\alpha_0$ and $\mu_{\alpha}\cos\delta$ taken from the posterior distribution of the standard fit, as described in \cite{SuppMaterial}.
\item[$^{\ddagger}$] Positions given for the reference date, MJD 56198.0.
\item[$^{\mathsection}$] Inclinations of 90--180\degr\ indicate clockwise orbits.
\end{tablenotes}
\end{threeparttable}
\end{table*}


\clearpage

\begin{thebibliography}{10}

\bibitem{Abbott2019}
B.~P. {Abbott}, {\it et~al.\/}, {\it \apjl\/} {\bf 882}, L24 (2019).

\bibitem{Casares2014}
J.~{Casares}, P.~G. {Jonker}, {\it \ssr\/} {\bf 183}, 223 (2014).

\bibitem{Orosz2007}
J.~A. {Orosz}, {\it et~al.\/}, {\it \nat\/} {\bf 449}, 872 (2007).

\bibitem{Belczynski2010}
K.~{Belczynski}, {\it et~al.\/}, {\it \apj\/} {\bf 714}, 1217 (2010).

\bibitem{Reid2011}
M.~J. {Reid}, {\it et~al.\/}, {\it \apj\/} {\bf 742}, 83 (2011).

\bibitem{Orosz2011}
J.~A. {Orosz}, {\it et~al.\/}, {\it \apj\/} {\bf 742}, 84 (2011).

\bibitem{Ziolkowski2014}
J.~{Zi{\'o}{\l}kowski}, {\it \mnras\/} {\bf 440}, L61 (2014).

\bibitem{Gaia2018}
{Gaia Collaboration}, {\it et~al.\/}, {\it \aap\/} {\bf 616}, A1 (2018).

\bibitem{Chan2020}
V.~C. {Chan}, J.~{Bovy}, {\it \mnras\/} {\bf 493}, 4367 (2020).

\bibitem{SuppMaterial}
Materials and methods are available as supplementary materials.

\bibitem{Brocksopp2002}
C.~{Brocksopp}, R.~P. {Fender}, G.~G. {Pooley}, {\it \mnras\/} {\bf 336}, 699
  (2002).

\bibitem{Astraatmadja2016}
T.~L. {Astraatmadja}, C.~A.~L. {Bailer-Jones}, {\it \apj\/} {\bf 832}, 137
  (2016).

\bibitem{Brocksopp1999a}
C.~{Brocksopp}, A.~E. {Tarasov}, V.~M. {Lyuty}, P.~{Roche}, {\it \aap\/} {\bf
  343}, 861 (1999).

\bibitem{Gies2003}
D.~R. {Gies}, {\it et~al.\/}, {\it \apj\/} {\bf 583}, 424 (2003).

\bibitem{Shimanskii2012}
V.~V. {Shimanskii}, {\it et~al.\/}, {\it Astronomy Reports\/} {\bf 56}, 741
  (2012).

\bibitem{Gou2011}
L.~{Gou}, {\it et~al.\/}, {\it \apj\/} {\bf 742}, 85 (2011).

\bibitem{Duro2016}
R.~{Duro}, {\it et~al.\/}, {\it \aap\/} {\bf 589}, A14 (2016).

\bibitem{Bressan2012}
A.~{Bressan}, {\it et~al.\/}, {\it \mnras\/} {\bf 427}, 127 (2012).

\bibitem{Russell:2007}
D.~M. {Russell}, R.~P. {Fender}, E.~{Gallo}, C.~R. {Kaiser}, {\it \mnras\/}
  {\bf 376}, 1341 (2007).

\bibitem{Kippenhahn1967}
R.~{Kippenhahn}, A.~{Weigert}, {\it \zap\/} {\bf 65}, 251 (1967).

\bibitem{Qin2019}
Y.~{Qin}, P.~{Marchant}, T.~{Fragos}, G.~{Meynet}, V.~{Kalogera}, {\it \apjl\/}
  {\bf 870}, L18 (2019).

\bibitem{Mirabel2003}
I.~F. {Mirabel}, I.~{Rodrigues}, {\it Science\/} {\bf 300}, 1119 (2003).

\bibitem{Grinberg2014}
V.~{Grinberg}, {\it et~al.\/}, {\it \aap\/} {\bf 565}, A1 (2014).

\bibitem{Tetarenko2016}
B.~E. {Tetarenko}, G.~R. {Sivakoff}, C.~O. {Heinke}, J.~C. {Gladstone}, {\it
  \apjs\/} {\bf 222}, 15 (2016).

\bibitem{Hurley2000}
J.~R. {Hurley}, O.~R. {Pols}, C.~A. {Tout}, {\it \mnras\/} {\bf 315}, 543
  (2000).

\bibitem{Vink2001}
J.~S. {Vink}, A.~{de Koter}, H.~J.~G.~L.~M. {Lamers}, {\it \aap\/} {\bf 369},
  574 (2001).

\bibitem{Arcavi:2017}
I.~{Arcavi}, {\it et~al.\/}, {\it \nat\/} {\bf 551}, 210 (2017).

\bibitem{Neijssel2019}
C.~J. {Neijssel}, {\it et~al.\/}, {\it \mnras\/} {\bf 490}, 3740 (2019).

\bibitem{Chruslinska2019}
M.~{Chruslinska}, G.~{Nelemans}, {\it \mnras\/} {\bf 488}, 5300 (2019).

\bibitem{MillerMiller:2015}
M.~C. {Miller}, J.~M. {Miller}, {\it \physrep\/} {\bf 548}, 1 (2015).

\bibitem{Farr2017}
W.~M. {Farr}, {\it et~al.\/}, {\it \nat\/} {\bf 548}, 426 (2017).

\bibitem{Miller-Jones2021}
J.~C.~A. {Miller-Jones}, {\it et~al.\/}, {\it Zenodo} {10.5281/zenodo.3961240} (2021).

\bibitem{Ma2009}
C.~{Ma}, {\it et~al.\/}, {\it IERS Technical Note\/} {\bf 35} (2009).

\bibitem{Pradel2006}
N.~{Pradel}, P.~{Charlot}, J.-F. {Lestrade}, {\it \aap\/} {\bf 452}, 1099
  (2006).

\bibitem{Condon1998}
J.~J. {Condon}, {\it et~al.\/}, {\it \aj\/} {\bf 115}, 1693 (1998).

\bibitem{Deller2011}
A.~T. {Deller}, {\it et~al.\/}, {\it \pasp\/} {\bf 123}, 275 (2011).

\bibitem{Greisen2003}
E.~W. {Greisen}, {\it Information Handling in Astronomy - Historical Vistas\/},
  A.~{Heck}, ed. (2003), vol. 285 of {\it Astrophysics and Space Science
  Library\/}, p. 109.

\bibitem{Reid2017}
M.~J. {Reid}, {\it et~al.\/}, {\it \aj\/} {\bf 154}, 63 (2017).

\bibitem{Charlot2020}
P.~{Charlot}, {\it et~al.\/}, {\it arXiv e-prints}, arXiv:2010.13625 (2020).

\bibitem{Szostek2007}
A.~{Szostek}, A.~A. {Zdziarski}, {\it \mnras\/} {\bf 375}, 793 (2007).

\bibitem{Tetarenko2019}
A.~J. {Tetarenko}, {\it et~al.\/}, {\it \mnras\/} {\bf 484}, 2987 (2019).

\bibitem{Yoon2015}
D.~{Yoon}, S.~{Heinz}, {\it \apj\/} {\bf 801}, 55 (2015).

\bibitem{Zanin2016}
R.~{Zanin}, {\it et~al.\/}, {\it \aap\/} {\bf 596}, A55 (2016).

\bibitem{Salvatier2016}
J.~Salvatier, T.~V. Wiecki, C.~Fonnesbeck, {\it {PeerJ} Computer Science\/}
  {\bf 2}, e55 (2016).

\bibitem{Neal2012}
R.~M. {Neal}, ``MCMC using Hamiltonian dynamics", in {\it Handbook of Markov Chain Monte Carlo}, S.~P. Brooks, A. Gelman, G.~L. Jones, X.-L. Meng, Eds. (Chapman and Hall/CRC Press, 2011), chap. 4

\bibitem{Betancourt2017}
M.~{Betancourt}, {\it arXiv e-prints\/} arXiv:1706.01520 (2017).

\bibitem{Hoffman2011}
M.~D. {Hoffman}, A.~{Gelman}, {\it ArXiv e-prints\/} p. arXiv:1111.4246 (2011).

\bibitem{Stirling2001}
A.~M. {Stirling}, {\it et~al.\/}, {\mnras\/} {\bf 327}, 1273 (2001).

\bibitem{Gaia2020}
{Gaia Collaboration}, {\it et~al.\/}, Gaia Early Data Release 3. Summary of the contents and survey properties
. {\it \aap\/}, 10.1051/0004-6361/202039657 (2021).

\bibitem{Lindegren2020}
L.~{Lindegren}, {\it et~al.\/}, {\it arXiv e-prints}, arXiv:2012.01742 (2020).

\bibitem{Luri2018}
X.~{Luri}, {\it et~al.\/}, {\it \aap\/} {\bf 616}, A9 (2018).

\bibitem{Rao2020}
A.~{Rao}, {\it et~al.\/}, {\it \mnras\/} {\bf 495}, 1491 (2020).

\bibitem{Gou2014}
L.~{Gou}, {\it et~al.\/}, {\it \apj\/} {\bf 790}, 29 (2014).

\bibitem{Blaauw1969}
A.~{Blaauw}, {\it \bain\/} {\bf 15}, 265 (1961).

\bibitem{Tomsick2014}
J.~A. {Tomsick}, {\it et~al.\/}, {\it \apj\/} {\bf 780}, 78 (2014).

\bibitem{Parker2015}
M.~L. {Parker}, {\it et~al.\/}, {\it \apj\/} {\bf 808}, 9 (2015).

\bibitem{Walton2016}
D.~J. {Walton}, {\it et~al.\/}, {\it \apj\/} {\bf 826}, 87 (2016).

\bibitem{Tomsick2018}
J.~A. {Tomsick}, {\it et~al.\/}, {\it \apj\/} {\bf 855}, 3 (2018).

\bibitem{Brocksopp1999b}
C.~{Brocksopp}, {\it et~al.\/}, {\it \mnras\/} {\bf 309}, 1063 (1999).

\bibitem{Caballero-Nieves2009}
S.~M. {Caballero-Nieves}, {\it et~al.\/}, {\it \apj\/} {\bf 701}, 1895 (2009).

\bibitem{Orosz2000}
J.~A. {Orosz}, P.~H. {Hauschildt}, {\it \aap\/} {\bf 364}, 265 (2000).

\bibitem{TerBraak2006}
C.~J.~F. {Ter Braak}, {\it Statistics and Computing\/} {\bf 16}, 239 (2006).

\bibitem{Herrero1995}
A.~{Herrero}, R.~P. {Kudritzki}, R.~{Gabler}, J.~M. {Vilchez}, A.~{Gabler},
  {\it \aap\/} {\bf 297}, 556 (1995).

\bibitem{Ziolkowski2005}
J.~{Zi{\'o}{\l}kowski}, {\it \mnras\/} {\bf 358}, 851 (2005).

\bibitem{McClintock2014}
J.~E. {McClintock}, R.~{Narayan}, J.~F. {Steiner}, {\it \ssr\/} {\bf 183}, 295
  (2014).

\bibitem{Novikov1973}
I.~D. {Novikov}, K.~S. {Thorne}, {\it Black Holes (Les Astres Occlus)\/}
  (1973), pp. 343--450.

\bibitem{Bardeen1972}
J.~M. {Bardeen}, W.~H. {Press}, S.~A. {Teukolsky}, {\it \apj\/} {\bf 178}, 347
  (1972).
  
\bibitem{McClintock2006}
J.~E. {McClintock}, {\it et~al.\/}, {\it \apj\/} {\bf 652}, 518 (2006).

  
\bibitem{Shafee2006}
R. {Shafee}, {\it et~al.\/}, {\it \apj\/} {\bf 636}, L113 (2006).

\bibitem{Steiner2009}
J.~F. {Steiner}, J.~E. {McClintock}, R.~A. {Remillard}, R.~{Narayan}, L.~{Gou},
  {\it \apjl\/} {\bf 701}, L83 (2009).

\bibitem{Thorne1974}
K.~S. {Thorne}, {\it \apj\/} {\bf 191}, 507 (1974).

\bibitem{Zhao2021}
X.~{Zhao}, {\it et~al.\/}, Reestimating the Spin Parameter of the Black Hole in Cygnus X-1, {\it \apj\/} {\bf accepted} (2021).

\bibitem{Higgins2019}
E.~R. {Higgins}, J.~S. {Vink}, {\it \aap\/} {\bf 622}, A50 (2019).

\bibitem{Puls2006}
J.~{Puls}, {\it et~al.\/}, {\it \aap\/} {\bf 454}, 625 (2006).

\bibitem{Puls:2008}
J.~{Puls}, J.~S. {Vink}, F.~{Najarro}, {\it \aapr\/} {\bf 16}, 209 (2008).

\bibitem{Vink:2017}
J.~S. {Vink}, {\it Philosophical Transactions of the Royal Society of London
  Series A\/} {\bf 375}, 20160269 (2017).


\bibitem{Humphreys1994}
R.~M. {Humphreys}, K.~{Davidson}, {\it \pasp\/} {\bf 106}, 1025 (1994).

\bibitem{Vink2005}
J.~S. Vink, A.~de~Koter, {\aap} {\bf 442}, 587 (2005).

\bibitem{Stevenson:2017}
S.~{Stevenson}, {\it et~al.\/}, {\it Nature Communications\/} {\bf 8}, 14906
  (2017).

\bibitem{VignaGomez:2018}
A.~{Vigna-G{\'o}mez}, {\it et~al.\/}, {\it \mnras\/} {\bf 481}, 4009 (2018).

\bibitem{Mennekens2014}
N.~{Mennekens}, D.~{Vanbeveren}, {\it \aap\/} {\bf 564}, A134 (2014).

\bibitem{Barrett:2017FIM}
J.~W. {Barrett}, {\it et~al.\/}, {\it \mnras\/} {\bf 477}, 4685 (2018).

\bibitem{Renzo:2017}
M.~{Renzo}, C.~D. {Ott}, S.~N. {Shore}, S.~E. {de Mink}, {\it \aap\/} {\bf
  603}, A118 (2017).

\bibitem{Maeder:1983}
A.~{Maeder}, {\it \aap\/} {\bf 120}, 113 (1983).

\bibitem{HaimanLoeb:1997}
Z.~{Haiman}, A.~{Loeb}, {\it \apj\/} {\bf 483}, 21 (1997).

\bibitem{Gotberg:2017}
Y.~{G{\"o}tberg}, S.~E. {de Mink}, J.~H. {Groh}, {\it \aap\/} {\bf 608}, A11
  (2017).

\bibitem{Prestwich2007}
A.~H. {Prestwich}, {\it et~al.\/}, {\it \apjl\/} {\bf 669}, L21 (2007).

\bibitem{Silverman2008}
J.~M. {Silverman}, A.~V. {Filippenko}, {\it \apjl\/} {\bf 678}, L17 (2008).

\bibitem{Carpano2007}
S.~{Carpano}, {\it et~al.\/}, {\it \aap\/} {\bf 466}, L17 (2007).

\bibitem{Crowther2010}
P.~A. {Crowther}, {\it et~al.\/}, {\it \mnras\/} {\bf 403}, L41 (2010).

\bibitem{Binder2011}
B.~{Binder}, {\it et~al.\/}, {\it \apj\/} {\bf 742}, 128 (2011).

\bibitem{Laycock2015}
S.~G.~T. {Laycock}, T.~J. {Maccarone}, D.~M. {Christodoulou}, {\it \mnras\/}
  {\bf 452}, L31 (2015).

\bibitem{Binder2015}
B.~{Binder}, J.~{Gross}, B.~F. {Williams}, D.~{Simons}, {\it \mnras\/} {\bf
  451}, 4471 (2015).

\bibitem{Liu2019}
J.~{Liu}, {\it et~al.\/}, {\it \nat\/} {\bf 575}, 618 (2019).

\bibitem{Shenar:2020}
T.~{Shenar}, {\it et~al.\/}, {\it \aap\/} {\bf 639}, L6 (2020).

\bibitem{AbdulMasih2019}
M.~{Abdul-Masih}, {\it et~al.\/}, {\it \nat\/} {\bf 580}, E11 (2020).

\bibitem{Eldridge2019}
J.~J. {Eldridge}, {\it et~al.\/}, {\it \mnras\/} {\bf 495}, 2786 (2020).

\bibitem{Elbadry2019}
K.~{El-Badry}, E.~{Quataert}, {\it \mnras\/} {\bf 493}, L22 (2020).

\bibitem{Kushnir:2016}
D.~{Kushnir}, M.~{Zaldarriaga}, J.~A. {Kollmeier}, R.~{Waldman}, {\it \mnras\/}
  {\bf 462}, 844 (2016).

\bibitem{Zaldarriaga:2017}
M.~{Zaldarriaga}, D.~{Kushnir}, J.~A. {Kollmeier}, {\it \mnras\/} {\bf 473},
  4174 (2018).

\bibitem{HotokezakaPiran:2017}
K.~{Hotokezaka}, T.~{Piran}, {\it \apj\/} {\bf 842}, 111 (2017).

\bibitem{Bavera:2019}
S.~S. {Bavera}, {\it et~al.\/}, {\it \aap\/} {\bf 635}, A97 (2020).

\bibitem{FullerMa:2019}
J.~{Fuller}, L.~{Ma}, {\it \apjl\/} {\bf 881}, L1 (2019).

\bibitem{Belczynski:2020}
K.~{Belczynski}, {\it et~al.\/}, {\it \aap\/} {\bf 636}, A104 (2020)

\bibitem{Axelsson:2011}
M.~{Axelsson}, R.~P. {Church}, M.~B. {Davies}, A.~J. {Levan}, F.~{Ryde}, {\it
  \mnras\/} {\bf 412}, 2260 (2011).

\bibitem{MandeldeMink:2016}
I.~{Mandel}, S.~E. {de Mink}, {\it \mnras\/} {\bf 458}, 2634 (2016).

\bibitem{Marchant:2016}
P.~{Marchant}, N.~{Langer}, P.~{Podsiadlowski}, T.~M. {Tauris}, T.~J. {Moriya},
  {\it \aap\/} {\bf 588}, A50 (2016).

\bibitem{Schroeder:2018}
S.~L. {Schr{\o}der}, A.~{Batta}, E.~{Ramirez-Ruiz}, {\it \apjl\/} {\bf 862}, L3
  (2018).

\bibitem{Podsiadlowski:2003}
P.~{Podsiadlowski}, S.~{Rappaport}, Z.~{Han}, {\it \mnras\/} {\bf 341}, 385
  (2003).

\bibitem{Valsecchi:2010}
F.~{Valsecchi}, {\it et~al.\/}, {\it \nat\/} {\bf 468}, 77 (2010).

\bibitem{Wong:2012}
T.-W. {Wong}, F.~{Valsecchi}, T.~{Fragos}, V.~{Kalogera}, {\it \apj\/} {\bf
  747}, 111 (2012).

\bibitem{Fryer:2012}
C.~L. {Fryer}, {\it et~al.\/}, {\it \apj\/} {\bf 749}, 91 (2012).

\bibitem{Israelian:1999}
G.~{Israelian}, R.~{Rebolo}, G.~{Basri}, J.~{Casares}, E.~L. {Mart{\'\i}n},
  {\it \nat\/} {\bf 401}, 142 (1999).

\bibitem{Podsiadlowski:2002SN}
P.~{Podsiadlowski}, {\it et~al.\/}, {\it \apj\/} {\bf 567}, 491 (2002).

\bibitem{Paxton:2011}
B.~{Paxton}, {\it et~al.\/}, {\it \apjs\/} {\bf 192}, 3 (2011).

\bibitem{Kippenhahn:1980}
R.~{Kippenhahn}, G.~{Ruschenplatt}, H.~C. {Thomas}, {\it \aap\/} {\bf 91}, 175
  (1980).

\bibitem{BraunLanger:1995}
H.~{Braun}, N.~{Langer}, {\it \aap\/} {\bf 297}, 483 (1995).

\bibitem{Neijssel:2020CygX1}
C.~{Neijssel}, {\it et~al.\/}, Wind mass-loss rates of stripped stars inferred
  from Cygnus~X-1, {\it \apj\/} {\bf accepted} (2021).
\end{thebibliography}
\end{document}